\documentclass[useAMS,usenatbib,usegraphicx,onecolumn]{mn2e}
\usepackage{bm}
\usepackage{times}

\title[Evolution of the grain size distribution]{What determines the grain size distribution in galaxies?}
\author[Asano et al.]{Ryosuke S. Asano$^{1}$\thanks{E-mail:
asano.ryosuke@g.mbox.nagoya-u.ac.jp}\thanks{Fellow of the Japan Society for the Promotion of Science (JSPS).},
Tsutomu T. Takeuchi$^{1}$,
Hiroyuki Hirashita$^{2}$,
and Takaya Nozawa$^{3}$ 
\\
$^{1}$Department of Particle and Astrophysical Science, Nagoya University, Furo-cho, Chikusa-ku, Nagoya 464-8602, Japan\\
$^{2}$Institute of Astronomy and Astrophysics, Academia Sinica, P.\ O.\
Box 23-141, Taipei 10617, Taiwan\\
$^{3}$Kavli Institute for the Physics and Mathematics of the Universe,
Todai Institutes for Advanced Study, the University of Tokyo, Kashiwa,
Chiba 277-8583, Japan}
\begin{document}

\date{Accepted 2013 March 18. Received 2013 March 18; in original form 2012 December 28}

\pagerange{\pageref{firstpage}--\pageref{lastpage}} \pubyear{2013}

\maketitle

\label{firstpage}

\begin{abstract}
Dust in galaxies forms and evolves by various processes, and 
these dust processes change the grain size distribution and amount of
 dust in the interstellar medium (ISM).
We construct a dust evolution model taking into account the grain size
 distribution, and investigate what kind of dust processes determine the grain
 size distribution at each stage of galaxy evolution.
In addition to the dust production by type II supernovae (SNe~II) and
 asymptotic giant branch (AGB) stars, we consider three processes in the ISM:
(i) dust destruction by SN shocks,
(ii) metal accretion onto the surface of preexisting grains in the
 cold neutral medium (CNM) (called grain growth), and (iii) grain{--}grain collisions (shattering and coagulation) in the warm neutral
 medium (WNM) and CNM.
We found that the grain size distribution in galaxies is controlled by
stellar sources in the early stage of galaxy evolution, 
and that afterwards the main processes that govern the size distribution
 changes to those in the ISM,
and this change occurs at earlier stage
 of galaxy evolution for a shorter star formation timescale (for star formation time-scales $= 0.5, 5$ and $50$~Gyr, the change
 occurs about galactic age $t \sim 0.6, 2$ and $5$~Gyr, respectively).
If we only take into account the processes which directly affect the total dust mass (dust production by SNe~II and AGB
 stars, dust destruction by SN shocks, and grain growth), the
 grain size distribution is biased to large grains ($a \sim
 0.2$--$0.5\;\mu$m, where $a$ is the grain radius).
Therefore, shattering is crucial to produce small ($a \la 0.01\;\mu$m) grains.
Since shattering produces a large abundance of small grains (consequently, the surface-to-volume ratio of grains increases), it
 enhances the efficiency of grain growth, contributing to the
 significant increase of the total dust mass.
Grain growth creates a large bump in the grain
 size distribution around $a \sim 0.01\;\mu$m.
Coagulation occurs effectively after the number of small grains is enhanced by shattering, and the grain size distribution is deformed to have a bump at
 $a \sim 0.03$--$0.05\;\mu$m at $t \sim 10$~Gyr.
We conclude that the evolutions of the total dust mass and the grain
 size distribution in galaxies are closely related to each other,
and the grain size distribution changes considerably through
the galaxy evolution because the dominant dust processes which regulate the grain
size distribution change.
\end{abstract}


\section{Introduction}
\label{sec:intro}

Dust is one of the most important factors for the understanding of galaxy
evolution.
Since hydrogen molecules are efficiently formed on the
surface of dust grains, the molecular formation rate is much larger than
the case without dust.
Such an enrichment of molecular abundance by dust realizes a favorable
condition for star formation \citep[e.g.,][]{hirashita02}. 
Dust grains also absorb stellar light mainly at ultraviolet and optical wavelengths and
re-emit in the infrared.
Consequently, dust affects the spectral energy distribution (SED) of
galaxies \citep[e.g.,][]{takagi}.
Furthermore, the formation rate of hydrogen
molecules on the grain surface and the mass absorption coefficient of radiation depend
strongly on the grain size distribution \citep[e.g.,][]{hirashita02,takeuchi03}.

Dust grains form by condensation of elements heavier than
helium (i.e., metals).
Metals are mainly supplied from asymptotic giant branch stars (AGB stars)
and supernovae (SNe), and part of them condense into dust grains
\citep[e.g,.][]{mathis}.
Dust grains are not only supplied by stars but are also destroyed
by SN shocks in the interstellar medium (ISM) \citep[e.g.,][]{jones96,nozawa06,zhukovska}.
Furthermore, it is thought that metal accretion onto the surface of grains
in the ISM (referred to as ``grain growth'' in this paper) is an important
process for explaining the amount of dust in the Milky Way
\citep[e.g.,][]{draine,pipino}.
To the present day, there have been a lot of studies that investigate the evolution
of the total dust mass in galaxies by taking into account these processes
\citep[e.g.,][]{dwek,dwek98,hirashita99a,hirashita99b,inoue03,inoue11,calura,zhukovska,pipino,asano}.
They assumed a representative grain size, but the efficiencies of dust destruction and grain growth depend
on the grain size distribution.
Thus, we should consider the evolution of the grain size distribution to
understand the total dust mass precisely.

The grain size distribution is derived from observed
extinction curves (which mainly depend on the grain size distribution and
the grain species).
According to \citet{MRN}, if spherical grains are assumed, the
extinction curve in the Milky Way is
reproduced by $f(a) {\rm d}a \propto a^{-3.5} {\rm d}a$
($0.005 < a < 0.25\;\mu$m; this grain size distribution is referred to
as the MRN distribution), where $a$ is the grain radius and $f(a)$d$a$ is
the number density of grains in size interval $[a,a+{\rm d}a]$
\citep[see][for more detailed fitting to the Milky Way extinction curve]{kim,WD01}.
The situation seems to be very different for distant galaxies. 
Recently, \citet{gallerani} discussed the extinction curves of seven quasars at high redshift ($3.9 \le z \le 6.4$).
They showed that these extinction curves tend to be flat at wavelengths
$< 0.2\;\mu$m in the quasar's rest frame.
The difference between extinction curves in distant and nearby objects
may indicate that different processes dominate the dust evolution
at different epochs.

In young galaxies, Type~II SNe (SNe~II) are thought to be the dominant sources of dust
because of short lifetime of their progenitors.
However, \citet{valiante09} suggested that AGB stars are also important
sources of dust production even at galactic age less than $1$~Gyr.
In addition, grain growth is expected to be the
dominant process to increase dust mass in galaxies if the metallicity becomes larger
than a certain value \citep{inoue11,asano}.
Furthermore, if the metallicity reaches a sub-solar value,
grain{--}grain collisions in the ISM (shattering and coagulation) become
efficient enough to change the grain size distribution significantly
\citep[e.g.,][]{hirashita09b}.
We call all processes affecting the grain size distribution `dust processes'.

These dust processes affect the different sizes of grains in galaxies.
\citet{nozawa07} showed that SNe~II supply relatively large
grains ($a \ga
0.01$~$\mu$m) into the ISM because small grains are destroyed by reverse
shocks before they are ejected into the ISM \citep[see also][]{bianchi,silvia}.
The size distribution of grains produced by AGB stars 
is thought to be biased to large ($\sim 0.1\;\mu$m) sizes
\citep[e.g.,][]{winters,groenewegen,yasuda}.
Furthermore, the smaller grains in the ISM are more easily destroyed by
interstellar shocks driven by SNe \citep[e.g.,][]{nozawa06}.
If grain growth occurs, since the timescale of this process is proportional to
the volume-to-surface ratio of a grain, smaller grains grow more efficiently \citep[e.g.,][]{hirashita11}.
After the dust grains are released into the diffuse ISM, shattering
can also occur.
\citet{yan} showed that large grains ($a \ga 0.1\;\mu$m) acquire larger
velocity dispersions than the shattering threshold velocity if the grains are
dynamically coupled with magnetized interstellar turbulence.
Shattering is indeed a promising mechanism of small{--}grain production \citep[e.g.,][]{hirashita10a}.
Shattering also occurs in SN shocks \citep{jones96}.
In dense regions, coagulation can occur,
so that the grain size distribution shifts to larger sizes \citep[e.g.,][]{ormel,hirashita09b}.
The various dust processes above in galaxies occur on timescales
dependent on the metallicity, the total dust amount, the grain size distribution, and
so on.
Hence, it is crucial to consider all dust processes in a unified
framework to understand the evolution of both the total dust amount and the grain size distribution.

There have been a number of studies on the evolution of the grain
size distribution in galaxies.
\citet{liffman} discussed the evolution of grain size distribution considering dust
destruction by SN shocks and grain growth.
However, they did not consider shattering and coagulation by
grain{--}grain collisions.
\citet{odonnell} suggested a dust evolution model in a multi-phase ISM
[warm neutral medium (WNM) and cold neutral medium (CNM)], and
also considered the collisional processes of dust grains.
However, they did not consider the size distribution of grains released by
stars in order to simplify their model.
\citet{hirashita10b} discussed the grain size distribution in young
starburst galaxies.
They assumed that SNe~II are the source of dust in
these galaxies and focused on the production of small grains by
shattering.
\citet{yamasawa} constructed a dust evolution model taking into account
dust formation and destruction by SNe~II along with the formation
and evolution of galaxies.
However, since they focused on galaxies in the high-$z$ Universe,
they did not consider dust formation in AGB stars, grain growth, shattering and coagulation.

In this work, we construct a dust evolution model taking into account
the dust formation by SNe~II and AGB stars, dust destruction by SN shocks, grain
growth, and shattering and coagulation, to investigate what kind of dust
processes determine the grain size distribution at each stage of galaxy
evolution.
In our model, we do not consider mass exchange among various ISM phases in
detail \citep[e.g.,][]{ikeuchi},
but our results contain the contributions of dust processes in the two
ISM phases, WNM ($\sim 6000$~K, $0.3\;{\rm cm}^{-3}$) and CNM ($\sim 100$~K,
$30\;{\rm cm}^{-3}$) by assuming these mass fractions in the ISM to be constant.

This paper is organized as follows:
in Section \ref{sec:model} we introduce the dust evolution model based on
chemical evolution of galaxies.
In Section \ref{sec:result} we examine the contribution of each dust process
to the grain size distribution.
Section \ref{sec:discussion} is devoted to the discussion on what kind of
dust processes regulate the grain size distribution in galaxies.
We conclude this work in Section \ref{sec:conclusion}.
Throughout this paper the solar metallicity is set to be $Z_{\odot} =
0.02$ \citep{anders}.

\section{Galaxy evolution model}
\label{sec:model}

In this Section, we introduce our dust evolution model in a galaxy.
First, we show the basic equations of the chemical evolution model. We then describe the
dust evolutions based on the chemical evolution model, involving dust production by SNe~II and AGB stars, dust
destruction by SN shocks, grain growth, and shattering and coagulation
by grain{--}grain collisions.

Some grain processing mechanisms work in a different way in a different
ISM phase \citep{odonnell}.
In this work, while we use a one-zone model to examine the representative
properties of a galaxy, we consider the effects of the dust processes
in WNM and CNM by introducing the mass fractions of WNM and CNM, $\eta_{\rm WNM}$
and $\eta_{\rm CNM}$.
Considering temperatures less than $10^4$~K,
we find that an equilibrium state of two thermally stable phases (WNM and CNM) is established in
the ISM \citep[e.g.,][]{wolfire}.
Thus, we calculate dust evolution taking into account
a two-phase neutral ISM.
We also assume that the galaxy is a closed-box; that is, the total
baryon mass $M_{\rm tot}$ (the sum of the stellar mass and the ISM mass in the galaxy) is constant.
Since $M_{\rm tot}$ is just a scale factor in our work, the total dust
mass just scales with $M_{\rm tot}$.
Throughout this paper $M_{\rm tot}$ is set to be $10^{10}\;\mbox{M}_{\odot}$.

Inflow and outflow are not considered in our model for simplicity.
Since inflowing gas is considered to be not only metal poor but also
dust poor, the abundance of both metals and dust are diluted with the
same (or similar) fraction by inflow.
This effect is degenerate with a slower chemical enrichment, under a longer $\tau_{\rm SF}$, where $\tau_{\rm
SF}$ is the star formation timescale. 
As for outflow, since ISM components (namely gas, metals, and dust) are blown out of
a galaxy, the total gas mass in a galaxy decreases.
In this case, star formation rate decreases at earlier phase of galaxy evolution;
that is, the effect of outflow is degenerate with a shorter $\tau_{\rm SF}$.
Thus, we just absorb the effects of inflow and outflow into $\tau_{\rm SF}$.

\subsection{Chemical evolution model}

In this subsection, we describe our model of chemical evolution in
a galaxy.
From the above assumptions, the equations of time evolution of the total stellar mass,
$M_*$, the ISM mass, $M_{\rm ISM}$, and the mass of a metal species X, $M_{\rm X}$, in the galaxy are expressed as
\begin{eqnarray}
\frac{{\rm d}M_*(t)}{{\rm d}t} &=& \mbox{SFR}(t) - R(t),\\
\frac{{\rm d}M_{\rm ISM}(t)}{{\rm d}t} &=& -\mbox{SFR}(t) + R(t),\\
\frac{{\rm d}M_{\rm X}(t)}{{\rm d}t} &=& -Z_{\rm X}(t)\mbox{SFR}(t) +
 Y_{\rm X}(t),
\end{eqnarray}
where $t$ is the galaxy age, $\mbox{SFR}(t)$ is the star formation rate, $Z_{\rm X} = M_{\rm
X}/M_{\rm ISM}$,
and $R(t)$ and $Y_{\rm X}(t)$ are the masses of the total baryons and
total metal species X released by
stars in a unit time, respectively.
In this paper, we consider two dust species, carbonaceous dust and
silicate dust, and we adopt
two key elements of dust species (X $=$ C for carbonaceous dust and X
$=$ Si for silicate dust) in calculating dust evolution (see Section
\ref{subsec:dustmodel} for details).
We adopt $M_*(0) = 0$, $M_{\rm ISM}(0) = M_{\rm tot}$, and $M_{\rm X}(0)
= 0$ as initial conditions.

In our work, we adopt the Schmidt law for the SFR:
$\mbox{SFR} \propto M^n_{\rm ISM}$ \citep{schmidt},
 and the index $n$ is thought to be $1${--}$2$ observationally
\citep[e.g.,][]{kennicutt}.
We here adopt $n = 1$,
\begin{equation}
\mbox{SFR}(t) = \frac{M_{\rm ISM}(t)}{\tau_{\rm SF}},
\label{eq:sfr}
\end{equation}
where the star formation timescale $\tau_{\rm SF}$ is a constant.
For comparison, the case with $n = 1.5$ is also shown in Appendix \ref{app:schmidt}.
As long as we adopt the same star formation timescale at $t = 0$, there is little difference between the two cases with $n = 1$ and $1.5$. 

$R(t)$ and $Y_{\rm X}(t)$ are written as
\begin{eqnarray}
R(t) &=& \int^{100\;{\rm M}_{\odot}}_{m_{\rm cut}(t)} \left[m -
						       \omega(m, Z(t -
						       \tau_m))\right] \phi(m)\mbox{SFR}(t){\rm d}m,
\label{eq:returnmass}\\
Y_{\rm X}(t) &=& \int^{100\;{\rm M}_{\odot}}_{m_{\rm cut}(t)} m_{\rm X}(m, Z(t -
 \tau_m)) \phi(m)\mbox{SFR}(t){\rm d}m,
\label{eq:returnmetal}
\end{eqnarray}
where $\phi(m)$ is the stellar initial mass function, $\tau_m$ is the lifetime
of a star with mass $m$ at the zero-age main sequence, $Z$ is the metallicity ($= \Sigma_X
M_{\rm X}/M_{\rm ISM}$), and 
$\omega(m,Z)$ and $m_{\rm X}(m,Z)$ represent the mass of remnant stars
(white dwarfs, neutron stars or black holes) and the
mass of metal species X ejected by a star of mass $m$ and
metallicity $Z$, respectively. 
For the lifetime of stars, we adopt the formula derived by
\citet{raiteri}, and the formula is obtained by the fitting to the
stellar models of the Padova group \citep{bertelli}.
Since its metallicity dependence is weak, we always adopt the stellar lifetime for solar metallicity as a
representative value.
The lower bound of the integration, $m_{\rm cut}(t)$ is the mass of a star with $\tau_m = t$.
We adopt the Salpeter IMF for stellar mass range
$0.1$--$100~\mbox{M}_{\odot}$ \citep{salpeter}:
\begin{equation}
\phi(m) \propto m^{-q},
\end{equation}
where $q$ is set to be $2.35$, and the normalization is determined by 
\begin{equation}
\int^{100\;{\rm M}_{\odot}}_{0.1\;{\rm M}_{\odot}} m\phi(m){\rm d}m = 1.
\end{equation}
To check the variation of the results with $q$, we examine the case with
$q = 1.35$ (a top heavy IMF) in Appendix \ref{app:imf}.
For $q = 1.35$, the processes in the ISM occur at earlier phases of
galaxy evolution than for $q = 2.35$, because a larger amount of dust is
supplied by stars.
However, we find that the sequence of the dominant dust processes
along the age does not change so the following discussions are not
altered significantly by the change of $q$.
Thus, we only consider $q = 2.35$ in the following discussion.

To calculate Eqs.~(\ref{eq:returnmass}) and (\ref{eq:returnmetal}), we
quote the remnant and metal mass data of
stars with mass $m$ and metallicity $Z$ from some previous works.
We assume that the mass ranges of AGB stars and SNe~II are
$1$--$8~\mbox{M}_{\odot}$ and $8$--$40~\mbox{M}_{\odot}$, respectively,
and that all stars with initial masses more than
$40~\mbox{M}_{\odot}$ evolve into black holes without ejecting any gas,
metals or dust \citep{heger}.
The data for AGB stars with mass
$1$--$6\;\mbox{M}_{\odot}$ and metallicity $Z = (0.005, 0.2, 0.4,
1.0)\;\mbox{Z}_{\odot}$ is taken from \citet{karakas} and the data for SNe~II with
mass $13$--$40\;\mbox{M}_{\odot}$ and metallicity $Z = (0.0, 0.05, 0.2,
1.0)\;\mbox{Z}_{\odot}$ is from \citet{kobayashi06}.
We interpolate and extrapolate the data for all values of mass and
metallicity (also for the dust data in Sections \ref{subsubsec:agb} and \ref{subsubsec:snII}).

\subsection{Dust evolution}
\label{subsec:dustmodel}

For dust evolution, we consider dust production by SNe~II and AGB stars, dust destruction by SN shocks in
the ISM, grain growth in the CNM,
and shattering and coagulation by grain{--}grain collisions in the WNM and CNM.
In this work, as mentioned in Section \ref{sec:intro}, we
assume a two-phase ISM (WNM and CNM) to
calculate the dust evolution (see also Section \ref{subsec:shattering}).

We neglect the contribution of Type Ia SNe (SNe~Ia) to the
production of metals and dust, and the destruction of dust.
\citet{nozawa11} showed that SNe~Ia release little dust into the ISM.
Furthermore, dust destruction by SNe~Ia is expected to be insignificant to the
total dust budget in galaxies (less than 1/10 of the contribution of
SNe~II; \citet{calura}).
As for metals, although \citet{nomoto} showed that the contribution of
SNe~Ia to the silicon and carbon enrichment in the ISM can be comparable
to that of SNe~II, the ratio between SN~Ia rate and SN~II rate is
unknown [\citet{nomoto} suggested that it is about $0.1$ taking into
account a chemical evolution model].
Thus, to simplify the discussion, we neglect the contribution
from SNe~Ia, keeping in mind a possible underproduction of metallicity.

 The dust production data we adopt contain a lot of dust species
 (C, Si, $\mbox{SiO}_2$, SiC, Fe, FeS, $\mbox{Al}_2\mbox{O}_3$, MgO,
 $\mbox{MgSiO}_3$, $\mbox{FeSiO}_3$, $\mbox{Mg}_2\mbox{SiO}_4$, and
 $\mbox{Fe}_2\mbox{SiO}_4$) \citep{nozawa07,zhukovska}.
 However, the physical properties of grain species other than
 carbonaceous and silicate grains are not fully known.
Hence, we categorize all grain species other than carbonaceous dust as silicate and calculate
 their growth, shattering, and coagulation by adopting the physical
 parameters of silicate grains.
In particular, after grain growth and coagulation occur, the
dust species categorized as silicate dust do not evolve separately and
our simplification can avoid the complexity arising from the compound
species.
In fact, the mass of dust grains ejected by SNe is dominated by Si
 grains, which would grow into silicate grains in the oxygen-rich
 environments such as molecular clouds. For carbonaceous dust, we adopt
 material properties of graphite. The adopted parameters of these two
 grain species are shown in Table \ref{tab:para} and are the same as in
 \citet{hirashita09b} and \citet{kuo}.
Although we calculate silicate and carbonaceous dust separately, we are
 interested in how the overall grain size distribution is affected by
 each dust process. Therefore, we focus on the total grain size distribution.
\begin{table*}
\centering
\begin{minipage}{140mm}
\caption{parameters for each dust species}
\begin{tabular}{lcccccccc}
\hline
Species&X&$g_{\rm X}$&
$m_{\rm X}$~[amu]&
s~$[\mbox{g~cm}^{-3}$]~${}^{\rm b}$
&
$v_{\rm shat}$~[km~$s^{-1}$]&
$\gamma~[\mbox{erg~cm}^{-2}]~{}^{\rm c}$
&
$E~[\mbox{dyn~cm}^{-2}]~{}^{\rm c}$
&
$\nu~{}^{\rm c}$
\\
\hline
Graphite&C&1.0&12&2.26&1.2&75&$1.0 \times 10^{11}$&0.32\\
Silicate&Si&0.166~${}^{\rm a}$&28.1&3.3&2.7&25&$5.4 \times 10^{11}$&0.17\\
\hline
\label{tab:para}
\end{tabular}
\vspace{-13pt}\\
 Note. X is the key element of dust species, $g_{\rm X}$ is the mass
 fraction of the key element X in the grains, $m_{\rm X}$ is the
 atomic mass of X, $s$ is the bulk density of dust grains, $v_{\rm
 shat}$ is the shattering threshold velocity, $\gamma$ is the surface
 energy per unit area of grains, $E$ is Young's modulus, and $\nu$ is
 Poisson's ratio.\\
\baselineskip=8pt\normalfont
${}^{\rm a}${\scriptsize We assume
 $\mbox{Mg}_{1.1}\mbox{Fe}_{0.9}\mbox{SiO}_4$ for the composition of
 silicate \citep{dralee}.}\\
${}^{\rm b}${\scriptsize \citet{dralee}.}\\
${}^{\rm c}${\scriptsize \citet{chokshi}.}\\
\end{minipage}
\end{table*}

In this work, we assume that grains are spherical.
Thus, the mass of a grain with radius $a$ is
\begin{equation}
m(a) = \frac{4 \pi a^3}{3}s,
\label{eq:particle}
\end{equation}
where $s$ is the bulk density of dust grains.
In our model, we consider that the minimum and maximum radii of
grains, $a_{\rm min}$ and $a_{\rm max}$, are $0.0003\;\mu$m and
$8\;\mu$m, respectively.
Although the minimum size of grains is poorly known,
even if $a_{\rm min} = 0.001\;\mu$m, 
the evolution of both the total dust mass and the grain size
distribution does not change significantly \citep{hirashita12a}.

\subsubsection{Dust production by AGB stars}
\label{subsubsec:agb}

The size distribution of grains produced by AGB stars is not well known. 
\citet{winters} suggested that the size
distribution is log-normal with a peak at $\sim 0.1\;\mu$m based on the
fitting to observed SEDs.
\citet{yasuda} have recently calculated the size distribution of
SiC produced by C-rich AGB stars by performing dust formation
calculation coupled with a hydrodynamical model.
They showed that the mass distribution, $a^4 f(a)$, is close to log-normal with a
peak at $0.2$--$0.3\;\mu$m, where the grain size distribution $f(a)$ is
defined so that $f(a){\rm d}a$ is the number density of dust grains
with radii in the range $[a, a + {\rm d}a]$ (The size distribution
multiplied by $a^4$ means the mass distribution per logarithmic grain
radius). 
Hence, both theory and observations suggest that AGB stars preferentially
produce large grains ($a \ga 0.1\;\mu$m).
In this paper, we simply assume that the mass distribution, $a^4 f(a)$, of
each species produced by AGB stars is log-normal with a peak at
$0.1\;\mu$m with standard deviation $\sigma = 0.47$, so that
the shape of the mass distribution in Fig.~7 in \citet{yasuda} is reproduced.
We normalize $f(a)$ by
\begin{equation}
m_{\rm d}(m) = \int^{\infty}_{0} \frac{4 \pi}{3} a^3 s f(a) {\rm d}a,
\end{equation}
where $m_{\rm d}(m)$ is the dust mass released by a star with mass $m$.
The size distributions of all species are assumed to be the same for simplicity.
Dust mass data for AGB stars with mass $1$--$7\;\mbox{M}_{\odot}$ and
metallicity $Z = (5.0 \times 10^{-2}, 0.1, 0.2,
0.4, 0.75, 1.0)\;\mbox{Z}_{\odot}$ is taken from \citet{zhukovska}.
The size distributions of dust species other than carbonaceous dust are
summed to compose the grain size distribution of silicate (the same
procedure is also applied in Section \ref{subsubsec:snII}).
We define $f_{\rm X}(a)$ as the size distribution of dust species, where
$X$ represents the key element of dust species (X = C for carbonaceous
dust and X = Si for silicate dust).

\subsubsection{Dust production by SNe~II}
\label{subsubsec:snII}

Some fraction of dust grains in galaxies are produced in the ejecta of SNe~II
\citep[e.g.,][]{matsuura}.
After a SN explosion, reverse shock occurs because of interactions between
the ISM surrounding the SN and its ejecta,
and dust grains are destroyed by sputtering in the shock \citep[e.g.,][]{bianchi,nozawa07}.
\citet{nozawa07} calculated the total mass and size distribution of dust
grains ejected by SNe~II considering the dust destruction in the radiative and non-radiative phases of SN
remnants.
We adopt the data for dust mass and size distribution
derived by \citet{nozawa07} for SNe~II with mass
$13$--$30\;\mbox{M}_{\odot}$ \footnote{Although \citet{nozawa07}
investigated only the dust formation in SNe~II evolving from
zero-metallicity stars, the grain species formed in the ejecta of SNe~II
and their size distribution are insensitive to the metallicity of
progenitor stars \citep[e.g.,][]{todini,kozasa}. In addition, the destruction
efficiiency of dust by the reverse shocks is almost independent of
metallicity in the ISM; its difference between $Z = 0$ and $Z =
Z_{\odot}$ is less than $15\%$ \citep[see][]{nozawa07}.}.
They also considered two cases for mixing in the helium core:
unmixed and mixed models.
\citet{hirashita05} showed that the data from the unmixed model is in better
agreement with observations than that of the mixed model.
Hence, we adopt the unmixed model.
\citet{nozawa07} showed that the size distribution of grains supplied by
SNe~II is biased to large ($\sim 0.1\;\mu$m) grains due to the destruction of small grains
by the reverse shock.

The amount and size distribution of grains injected by SNe~II depend on the
density of the surrounding ISM because the dust destruction efficiency of the reverse shock is higher in
the denser ISM.
However, the trend that smaller grains are more easily destroyed does
not change.
In this paper, the hydrogen number density of the ISM surrounding the SNe~II, $n_{\rm
SN}$, is set to be $1.0\;{\rm cm}^{-3}$ as a fiducial value, but the
cases with $n_{\rm SN} = 0.1$ and $10.0\;{\rm cm}^{-3}$ are also examined.

\subsubsection{Dust destruction by SN shocks in the ISM}

Dust grains in the ISM are destroyed or become smaller by sputtering due
to the passage of interstellar shocks driven by SNe.
Since the destruction changes and depends on the grain size,
it is important to consider dust destruction taking into
account the grain size distribution.

To calculate this destruction process, we adopt the formulae in \citet{yamasawa}, which
we show here briefly.
The number density of dust grains with radii in the range $[a, a +
{\rm d}a]$ after the passage of a SN shock, $f'_{\rm X}(a){\rm d}a$, is given by 
\begin{equation}
f'_{\rm X}(a){\rm d}a = \int^{\infty}_{a} \xi_{\rm X}(a,a'){\rm d}a f_{\rm X}(a'){\rm d}a',
\end{equation}
where $\xi_{\rm X}(a,a'){\rm d}a$ is the number fraction of grains that
are eroded from the initial radii $[a', a' + {\rm d}a']$ to radii $[a, a +
{\rm d}a]$ by sputtering in the SN shock and has been obtained using the
models by \citet{nozawa06}. 
Note that if $a > a'$, $\xi_{\rm X}(a,a') = 0$.
Thus, the change in the number density of grains with radii $[a, a + {\rm d}a]$,
 ${\rm d}N_{\rm d, X}(a)$, after the passage of a single SN shock is expressed
as 
\begin{equation}
{\rm d}N_{\rm d, X}(a) = \int^{\infty}_0 \xi_{\rm X}(a,a'){\rm d}a f_{\rm
 X}(a'){\rm d}a' - f_{\rm X}(a){\rm d}a.
\end{equation}
Accordingly, the change of mass density, ${\rm d}M_{\rm d, X}(a)$, is 
\begin{eqnarray}
\nonumber{\rm d}M_{\rm d, X}(a) &=& \frac{4}{3}\pi a^3 s {\rm d}N_{\rm d, X}(a)\\
&=& \int^{\infty}_0 \frac{4 \pi a^3 s}{3} \xi_{\rm
 X}(a,a'){\rm d}a f_{\rm X}(a'){\rm d}a' - {\cal M}_{\rm d, X}(a){\rm d}a,
\end{eqnarray}
where ${\cal M}_{\rm d, X}(a){\rm d}a = \frac{4}{3}\pi a^3 s f_{\rm X}(a) {\rm d}a
= M_{\rm d, X}(a)$ is the total dust mass with radii $[a, a +
{\rm d}a]$ before the dust destruction.
The dust destruction efficiency $\xi_{\rm X}$ depends on the hydrogen
number density of the ISM, $n_{\rm SN}$, such that dust grains are destroyed
more efficiently in denser regions.
As mentioned in Section \ref{subsubsec:snII}, $n_{\rm SN} = 1.0\;{\rm cm}^{-3}$ as a fiducial
value in this paper.

The equation for the time evolution of $M_{\rm d, X}(a,t)$ for dust destruction by SN shocks in the ISM is expressed as 
\begin{equation}
\frac{{\rm d}M_{\rm d, X}(a,t)}{{\rm d}t} = -\frac{M_{\rm swept}}{M_{\rm
  ISM}(t)} \gamma_{\rm SN}(t)\left[M_{\rm d, X}(a,t)-m(a)\int^{\infty}_{0}\xi_{\rm X}(a,a^{'}){\rm
			     d}a f_{\rm X}(a^{'},t){\rm d}a'\right],
\label{eq:sndest}
\end{equation}
where $\gamma_{\rm SN}(t)$ is the SN rate and $M_{\rm swept}$ is the ISM
mass swept up by a SN shock. 
To express the dependence on the galaxy age, we write $M_{\rm d, X}(a)$ and $f_{\rm
X}(a)$ as $M_{\rm d, X}(a,t)$ and $f_{\rm X}(a,t)$, respectively.

The SN rate, $\gamma_{\rm SN}(t)$, is expressed as
\begin{equation}
\gamma_{\rm SN}(t) = \int^{40\;{\rm M}_{\odot}}_{8\;{\rm
 M}_{\odot}}\phi(m)\mbox{SFR}(t-\tau_m){\rm d}m,
\label{eq:snrate}
\end{equation}
where we assume that the range of integration in Eq.~(\ref{eq:snrate}) is the mass
range where SNe occur \citep{heger} (if $t - \tau_m < 0$, $\mbox{SFR}(t-\tau_m) = 0$).

The ISM mass swept up by a SN shock, $M_{\rm swept}$, depends on the density and
metallicity of the ISM.
In our model, we adopt the following formula used in \citet{yamasawa};
\begin{equation}
\frac{M_{\rm swept}}{M_{\odot}} = 1535n^{-0.202}_{\rm SN}\left[\left( \frac{Z}{
Z_{\odot}}\right) + 0.039\right]^{-0.289}.
\end{equation}

\subsubsection{Grain growth}

Here, we formulate the growth process of grains taking
into account the grain size distribution.
In the ISM, particularly in dense and cold regions, metals
accrete onto the surface of grains efficiently \citep[e.g.,][]{liffman,inoue03,draine}.
Recently, various studies have shown the importance of grain
growth for dust enrichment in galaxies \citep[e.g.,][]{zhukovska,michalowski,pipino,valiante11,hirashita11}.
\citet{hirashita11} showed quantitatively that the grain
size distribution has a very important consequence for grain
growth.
Here, we follow the formulation by \citet{hirashita11} and consider only
grain growth of refractory dust (silicate and carbonaceous dust in this paper).
Although volatile grains such as water ice also exists in clouds in reality, they
evaporate quickly when the clouds disappear or the gas temperature rises.

For grain growth, the following equation holds:
\begin{equation}
\frac{\partial f_{\rm X}(a,t)}{\partial t} + \frac{\partial}{\partial
 a}[f_{\rm X}(a,t)\dot{a}] = 0,
\label{eq:growth}
\end{equation}
where $\dot{a} \equiv {\rm d}a/{\rm d}t$ is the growth rate of the grain
radius \footnote{Note that Eq.~(\ref{eq:growth}) is valid for the case
where only grain growth is considered, i.e., without sputtering, shattering and coagulation.}.

From Eq.~(\ref{eq:particle}),
\begin{equation}
\frac{{\rm d}m(a)}{{\rm d}a} = 4 \pi a^2 s.
\label{eq:sizediff}
\end{equation}
Also, from \citet{hirashita11}, the rate of mass increase of a grain with radius
$a$ is expressed as
\begin{equation}
\frac{{\rm d}m(a)}{{\rm d}t} = g^{-1}_{\rm X} m_{\rm X} \alpha {\cal R},
\end{equation}
where $g_{\rm X}$ is the mass fraction of the key species X in the
grains, $m_{\rm X}$ is the atomic mass of X, $\alpha$ is the sticking
coefficient of the key species, and ${\cal R}$ is the collision rate of X to a grain with
radius $a$, defined as follows \citep{evans}
\begin{equation}
{\cal R} = 4 \pi a^2 n_{\rm X}(t) \left(\frac{kT_{\rm gas}}{2 \pi m_{\rm X}}\right)^{1/2},
\end{equation}
where $n_{\rm X}(t)$ is the number density of X in the gas phase
in the CNM, $k$ is the Boltzmann constant,
and $T_{\rm gas}$ is the gas temperature in the CNM.
We apply $g_{\rm X} = 1.0$ and $0.166$ for carbonaceous and silicate
dust, respectively (Table $1$) and $T_{\rm gas} = 100$~K.

Next, we consider $n_{\rm X}(t)$, which is evaluated by
\begin{equation}
n_{\rm X}(t) = \frac{\rho^{\rm eff}_{\rm ISM}}{m_{\rm X}} \frac{M_{\rm X}(t)}{M_{\rm
 ISM}(t)} \left[1 - g_{\rm X} \frac{M_{\rm d,X}(t)}{M_{\rm
	   X}(t)}\right],
\label{eq:numchange}
\end{equation}
where $\rho^{\rm eff}_{\rm ISM}$ is the average mass density of the ISM in
which grain growth occurs.
As grain growth occurs, the number of gaseous metals decreases.
Thus, $n_{\rm X}$ is a decreasing function of time if only grain growth
is concerned.
The mass density is estimated as $\rho^{\rm eff}_{\rm ISM} = \mu
m_{\rm H} n_{\rm H, CNM}$, where $\mu$ is the mean atomic weight,
assumed to be $1.33$ (the mass ratio of
hydrogen to helium is $3:1$).
In addition, $m_{\rm H}$ and $n_{\rm H, CNM}$ are the mass of a hydrogen
atom and the hydrogen number density in the CNM, respectively, and 
we apply $n_{\rm CNM} = 30\;{\rm cm}^{-3}$.
Hence, from the above four equations
[Eq.~(\ref{eq:sizediff})--(\ref{eq:numchange})], we obtain
\begin{equation}
\dot{a} \equiv \frac{{\rm d}a}{{\rm d}t} = \frac{\alpha \rho^{\rm
 eff}_{\rm ISM}}{g_{\rm X}s} \frac{M_{\rm X}(t)}{M_{\rm
 ISM}(t)} \left(\frac{kT_{\rm gas}}{2 \pi m_{\rm X}}\right)^{1/2}
 \left[1 - g_{\rm X} \frac{M_{\rm d,X}(t)}{M_{\rm X}(t)}\right].
\label{eq:radiusgrowth}
\end{equation}
We assume $\alpha = 1$ for simplicity, which means that when the key species collide with a dust grain, it definitely sticks.
In our study, we calculate the grain growth using
Eqs.~(\ref{eq:growth}) and (\ref{eq:radiusgrowth}).

\subsubsection{Shattering}
\label{subsec:shattering}

Turbulence occurs in the ISM ubiquitously,
and it is confirmed that turbulence is
maintained by thermal conduction from simulations \citep[e.g.,][]{koyama}.
In a turbulent medium, dust grains are accelerated by turbulence
\citep[e.g.,][]{mckee}, and they collide with each other and shattering
can occur \citep[e.g.,][]{yan,hirashita09b,hirashita10b}.
\citet{hirashita09b} suggested that the grain size distribution in the
ISM changes significantly by shattering due to collisions between dust grains
accelerated by magnetohydrodynamic turbulence \citep{yan}.
In our model, to calculate shattering process, we adopt the grain
velocity calculated by \citet{yan}, and the shattering equation and parameters 
used by \citet{hirashita09b}, whose formulation is based on \citet{jones96}.

We outline the treatment of shattering.
We define $\rho_{\rm X}(a,t){\rm d}a = m(a) f_{\rm X}(a,t){\rm d}a$ as
the mass of grains with radii [$a,a+{\rm d}a$] in a unit volume (refer
to as ``mass density'' in this paper).
Considering shattering in the collision between two grains with
radii $a_1$ and $a_2$ (called grains $1$ and $2$, respectively), the time evolution of $\rho_{\rm X}(a,t){\rm d}a$ for shattering is
expressed as
\begin{eqnarray}
\nonumber\hspace{-3mm}\left[\frac{{\rm d}\rho_{\rm X}(a,t){\rm d}a}{{\rm d}t}\right]_{\rm shat}\hspace{-6mm}
 = &-&m(a)\rho_{\rm X}(a,t){\rm d}a\!\!\int^{a_{\rm max}}_{a_{\rm
min}}\hspace{-3mm}\alpha[m(a),m(a_1)]\rho_{\rm X}(a_{1},t){\rm d}a_{1}
{\rm d}a_{1}\!\\
&+&\!\int^{a_{\rm max}}_{a_{\rm min}}\!\!\!\int^{a_{\rm
  max}}_{a_{\rm min}}\hspace{-3mm}\alpha[m(a_{1}),m(a_{2})]\rho_{\rm X}(a_{1},t){\rm
  d}a_{1}\rho_{\rm X}(a_{2},t){\rm d}a_{2} m^{1,2}_{\rm shat}(a) {\rm d}a_{1}{\rm
  d}a_{2},
\label{eq:shattering}
\end{eqnarray}
and
\begin{eqnarray}
\alpha[m(a_{1}),m(a_{2})] = \left\{
  \begin{array}{ll}
{\displaystyle   0} & (v_{1,2} \le v_{\rm shat})\\
   \frac{\displaystyle \sigma_{1,2}v_{1,2}}{\displaystyle m(a_{1})m(a_{2})} & (v_{1,2} > v_{\rm shat}),
  \end{array}
\right.
\label{eq:shatteredrate}
\end{eqnarray}
where $m^{1,2}_{\rm shat}(a)$ is the total mass of shattered fragments
of grain $1$ within size bin $[a, a + {\rm d}a]$ by a
collision between grains $1$ and $2$, and depends on the relative velocity of the grains.
The size distribution of shattered fragments is proportional to
$a^{-3.3}$ \citep[e.g.,][]{jones96} \footnote{The method of calculation of
the maximum and minimum size of fragments is described in detail in
Section 2.3 in
\citet{hirashita09b}.}.
The cross section of the collision between grains $1$ and $2$ is assumed
to be $\sigma_{1,2} =\pi (a_{1} + a_{2})^2]$.
The shattering threshold, $v_{\rm shat}$, is assumed to be $1.2~{\rm
km~s}^{-1}$, and $2.7~{\rm km~s}^{-1}$ for carbonaceous dust and
silicate dust, respectively \citep{jones96}.
We adopt the same treatment for the relative velocity as \citet{jones94}
and \citet{hirashita09b}: Each time step is divided into four small time steps,
and we consider shattering under the following four relative velocities in each small
time step (i) front collision $(v_{1,2} = v_1 + v_{2})$,
(ii) back-end collision $(v_{1,2} = |v_{1} - v_{2}|)$,
(iii) side collision $v_{1,2} = v_{1}$,
and (iv) $v_{1,2} = v_{2}$.
Here, $v_{1}$ and $v_{2}$ are the velocities of the grain with radius
$a_1$ and $a_{2}$, respectively.

Shattering can occur not only in turbulence but also in
SN shocks \citep[e.g.,][]{jones96}.
However, both of these shattering mechanisms have similar consequences on
the grain size distribution, so it is difficult to separate them.
To compare our work with previous studies \citep{hirashita10b,kuo},
we only consider shattering in turbulence.

\subsubsection{Coagulation}

In low temperature and high density regions of the ISM, 
it is expected that coagulation by grain{--}grain collisions occurs.
Indeed, \citet{stepnik} observed dense filaments and showed
that the ratio of the intensity in the filaments, $I_{60\;\mu{\rm m}}/I_{100\;\mu{\rm m}}$, is
smaller than that in the diffuse ISM.
They concluded that this trend resulted from the decrease of small
grains due to coagulation.
For coagulation, we adopt the formulation, the velocity of grains, and the parameters used by \citet{hirashita09b}.

The time evolution of $\rho_{\rm X}(a,t){\rm d}a$ for coagulation is
expressed as follows
\begin{eqnarray}
\nonumber\hspace{-3mm}\left[\frac{{\rm d}\rho_{\rm X}(a,t){\rm d}a}{{\rm
	      d}t}\right]_{\rm coag}\hspace{-6mm}=&-&m(a)\rho_{\rm X}(a,t){\rm d}a\!\!\int^{a_{\rm max}}_{a_{\rm
 min}}\hspace{-3mm}\alpha[m(a_{1}),m(a)]\rho_{\rm X}(a_{1},t){\rm d}a_{1}{\rm
 d}a_{1}\!\\
&+&\!\int^{a_{\rm max}}_{a_{\rm min}}\!\!\!\!\int^{a_{\rm
 max}}_{a_{\rm min}}\hspace{-3mm}\alpha[m(a_{1}),m(a_{2})]\rho_{\rm X}(a_{1},t){\rm
  d}a_{1}\rho_{\rm X}(a_{2},t){\rm d}a_{2}m^{1,2}_{\rm coag}(a){\rm d}a_{1}{\rm
  d}a_{2},
\end{eqnarray}
and 
\begin{eqnarray}
\alpha[m(a_{1}),m(a_{2})] = \left\{
  \begin{array}{ll}
  {\displaystyle 0 }& (v_{1,2} \ge v^{1,2}_{\rm coag})\\
   \frac{\displaystyle \beta\sigma_{1,2}v_{1,2}}{\displaystyle m(a_{1})m(a_{2})} & (v_{1,2} < v^{1,2}_{\rm coag}),
  \end{array}
\right.
\label{eq:coagthreshold}
\end{eqnarray}
where $\beta$ is the sticking coefficient of dust grains, and
$m^{1,2}_{\rm coag}(a) = m(a_1)$ if the mass range of $m(a_{1})+m(a_{2})$ is
within $[m(a),m(a)+{\rm d}m(a)]$; otherwise $m^{1,2}_{\rm coag}(a) = 0$.

We assume that coagulation occurs if the relative velocity is less than
the coagulation threshold $v^{1,2}_{\rm coag}$.
In our model, it is calculated in the same way as \citet{hirashita09b},
\begin{equation}
v^{1,2}_{\rm coag} = 21.4 \left[\frac{a^{3}_{1} +
					a^{3}_{2}}{(a_{1} +
					a_{2})^3}\right]^{1/2}
\frac{\gamma^{5/6}}{E^{1/3} R^{5/6}_{1,2} s^{1/2}},
\label{eq:coagthrevel}
\end{equation}
where $R_{1,2} \equiv a_1 a_2/(a_1 + a_2)$, $\gamma$ is the surface
energy per unit area, and $1/E = [(1 - \nu_1)^2/E_1 + (1 - \nu_2)^2/E_2]$,
where $\nu_1$ and $E_1$ are Poisson's ratio and Young's modulus of grain
$1$.
The parameters we used are shown in Table \ref{tab:para}.
Here, we assume $\beta = 1$ for simplicity.
The treatment of the relative velocity is the same as for shattering.

\subsubsection{Formulation of the grain-size dependent evolution of dust mass}
\label{subsubsec:formula}

Here, using the dust processes introduced above, we show the equation for
the dust mass evolution in a galaxy at each grain radius bin, so that we
can finally obtain the evolution of grain size distribution.
Defining $\Delta M_{\rm d}(a,t) \equiv m(a)f(a,t)\Delta a$
as the mass density of grains with radii $[a, a+\Delta a]$
\footnote{In this Section, we use the symbol ``$\Delta$'' to
emphasize that it stands not for infinitesimal but a certain small amount.} and a
galactic age $t$, it is formulated as
\begin{eqnarray}
\nonumber \frac{{\rm d} \Delta M_{\rm d}(a,t)}{{\rm d}t} &=& -\frac{\Delta M_{\rm d}(a,t)}{M_{\rm
 ISM}(t)} + \Delta Y_{\rm d}(a,t)\\
 \nonumber &&\hspace{-9pt} -\frac{M_{\rm swept}}{M_{\rm ISM}(t)}\gamma_{\rm SN}(t)\left[\Delta M_{\rm
			     d}(a,t)-m(a)\int^{\infty}_{0}\xi(a,a^{'})\Delta
			     a f(a^{'},t){\rm d}a'\right]\\
 \nonumber &&\hspace{-9pt} +\eta_{\rm CNM} \left[m(a) \Delta a\frac{\partial [f(a,t)]}{\partial t}\right]\\
 \nonumber &&\hspace{-9pt} +\eta_{\rm WNM}\left[\frac{{\rm d}\Delta M_{\rm d}(a,t)}{{\rm
		      d}t}\right]_{\rm shat, WNM} + \eta_{\rm
 CNM}\left[\frac{{\rm d}\Delta M_{\rm d}(a,t)}{{\rm d}t}\right]_{\rm shat, CNM}\\
  && \hspace{-9pt} +\eta_{\rm WNM}\left[\frac{{\rm d}\Delta M_{\rm d}(a,t)}{{\rm
		    d}t}\right]_{\rm coag, WNM} + \eta_{\rm
 CNM}\left[\frac{{\rm d}\Delta M_{\rm d}(a,t)}{{\rm
		    d}t}\right]_{\rm coag, CNM},
\label{eq:dustevo}
\end{eqnarray}
where $\eta_{\rm WNM}$ and $\eta_{\rm CNM}$ are the mass fraction of WNM
and CNM in the ISM, respectively.
From top to bottom, the terms on the right hand side describe reduction of dust due to astration, ejection of dust from stellar
sources, dust destruction by SN shocks,
grain growth in the CNM,
shattering in the WNM and CNM,
and coagulation in the WNM and CNM.
To calculate the dust processes which occur in each ISM
phase, as mentioned in Section \ref{sec:intro}, we assume 
(1) that $\eta_{\rm WNM}$ and $\eta_{\rm CNM}$ are constants and
(2) that there are two stable phases, WNM and CNM, in the ISM (namely, the sum of $\eta_{\rm WNM}$ and $\eta_{\rm CNM}$ is unity).

The total mass of grains with radii $[a,a+\Delta a]$ ejected by stars
per unit time, $\Delta Y_{\rm d}(a,t)$, 
is expressed as
\begin{equation}
\Delta Y_{\rm d}(a,t) = \int^{100\;{\rm M}_{\odot}}_{m_{\rm cut}(t)}
 \Delta m_{\rm d}(m,Z(t - \tau_m),a) \phi(m)\mbox{SFR}(t-\tau_m){\rm
 d}m,
\label{eq:stardust}
\end{equation}
where $\Delta m_{\rm d}(m,Z,a)$ is the total mass of grains with radii
$[a,a+\Delta a]$ released by stars with mass $m$ and metallicity $Z$.

\section{Model results}
\label{sec:result}

In this paper, as mentioned above, we consider the effects of dust
formation by SNe~II and AGB stars, dust
destruction by SN shocks in the ISM, grain growth, shattering, and
coagulation on the evolution of grain size distribution in galaxies.
Among these processes, dust formation by SNe~II and AGB stars, dust destruction and grain growth
directly increase or decrease the total dust mass,
while shattering and coagulation modify only the grain size
distribution. 
The evolution of the total dust mass in galaxies is often modeled by taking into account
the former four contributions (dust formation by SNe~II and AGB stars,
dust destruction, and grain growth) \citep[e.g.,][]{dwek,dwek98,calura,zhukovska,pipino,inoue11,asano}.
They calculated the dust evolution by assuming a representative grain
size, but the dust destruction and grain growth depend
on the grain size distribution. 
Thus, it is unknown whether these four contributions can reproduce the grain
size distribution in nearby galaxies even though they can explain the evolution of the
total dust mass.
In Section \ref{subsec:w/o collision}, we first investigate the
contributions of the processes that directly affect the total dust mass,
and then in Section \ref{subsec:w/ collision}, we examine the effects
of shattering and coagulation.

\subsection{Without the effects of grain{--}grain collisions}
\label{subsec:w/o collision}

\subsubsection{Stellar processes}
\label{subsubsec:stellar}

\begin{figure}
\centering\includegraphics[width=0.45\textwidth]{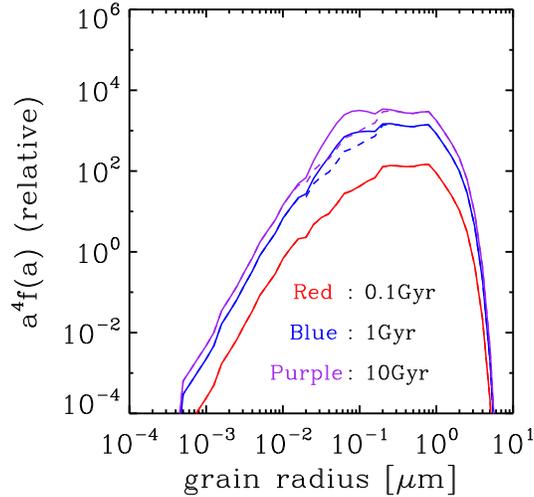}
\caption{Grain size distribution taking into account the dust production
 by AGB stars and SNe~II and dust reduction through astration (solid lines). Red,
 blue, and purple lines represent the cases at $t = 0.1, 1.0$,
 and $10$~Gyr, respectively, with $\tau_{\rm SF} = 5$~Gyr and $n_{\rm SN} =
 1.0\;{\rm cm}^{-3}$.
Dashed lines are cases with dust production by SNe~II only and dust
 reduction through astration, the same color
 corresponding to the same age. Note that the red dashed line overlaps
 with the red solid line.}
\label{fig:staronly}
\end{figure}

First, we consider the stellar processes including dust ejection from
stars (SNe~II and AGB stars) and dust reduction via astration.
Figure \ref{fig:staronly} shows the result.
The size distribution is expressed by multiplying $a^4$ to show the
mass distribution in logarithmic grain radius bin.
We adopt $\tau_{\rm SF} = 5$~Gyr and $n_{\rm SN} = 1.0\;{\rm cm}^{-3}$.
We also show the cases with SNe~II only.
As mentioned in Section \ref{sec:model}, since $M_{\rm tot}$ is just a
scale factor, the shape of the size distribution does not change with
$M_{\rm tot}$.
From Fig.~\ref{fig:staronly}, throughout any galactic age, we can observe that the grain size
distribution has a peak at around $0.5~\mu$m, and that only a small amount
of grains with $a < 0.01~\mu$m can be formed by stars.
\begin{figure*}
\centering\includegraphics[width=0.45\textwidth]{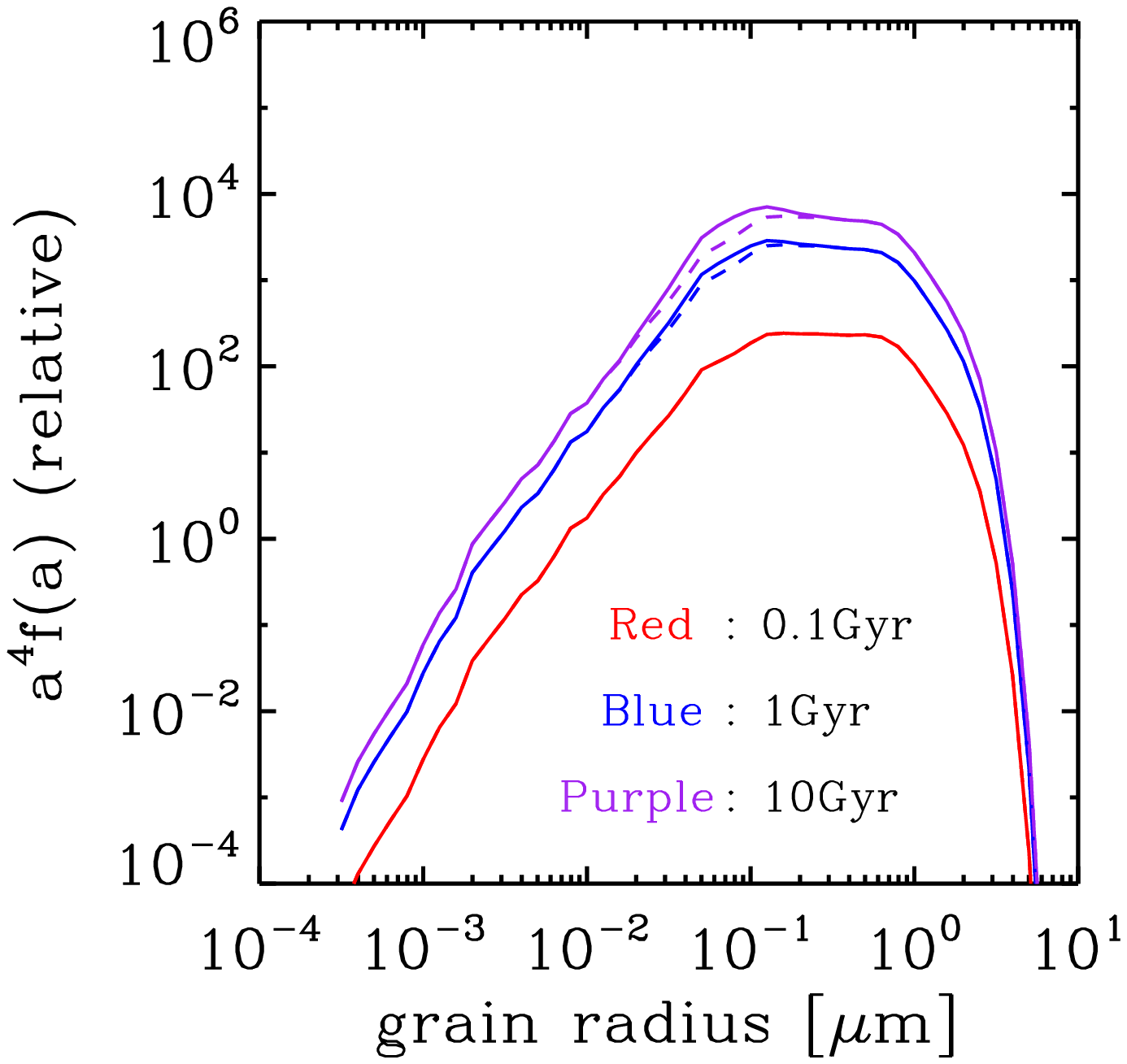}
\includegraphics[width=0.45\textwidth]{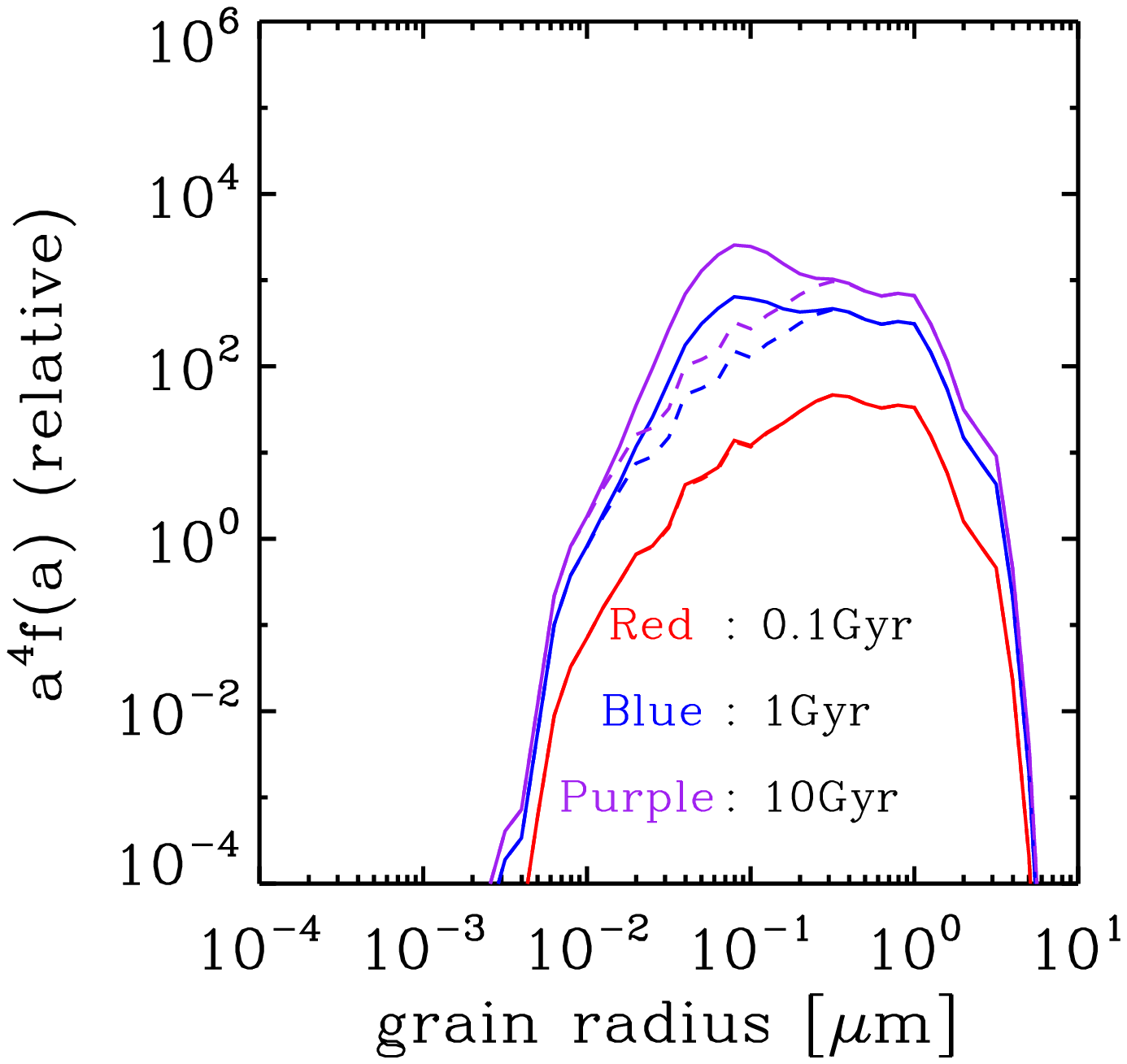}
\caption{Same as in Fig.~\ref{fig:staronly}, but we adopt different
 values of $n_{\rm SN}$: $0.1~{\rm cm}^{-3}$ in the left panel and $10.0~{\rm
 cm}^{-3}$ in the right panel.
Note that the red dashed lines overlap with the
 red solid lines.}
\label{fig:star_chansn}
\end{figure*}
In Fig.~\ref{fig:star_chansn}, we show the grain size distribution for other values of
$n_{\rm SN}$: $0.1~{\rm cm}^{-3}$ in the left panel and $10.0~{\rm
 cm}^{-3}$ in the right panel, respectively.
From Figure~\ref{fig:star_chansn}, we find that a larger amount of dust
grains with radii less than $\sim 0.1~\mu$m are destroyed by reverse
shocks in the case of higher $n_{\rm SN}$, and a smaller amount of
dust is supplied to the ISM.
However, even if $n_{\rm SN}$ changes, the trend that a
small amount of dust grains with radii less than
$0.01~\mu$m are supplied to the ISM does not change.
Thus, stars are the sources of dust grains with large radii ($\ge 0.05\;\mu$m). 

From Figs.~\ref{fig:staronly} and \ref{fig:star_chansn}, we observe that
dust from SNe~II always dominates the grain size distribution, while the
contribution of dust from AGB stars is seen only around
$0.1\;\mu$m at a galactic age $t = 10$~Gyr.
From our calculation, for the case with $n_{\rm SN} = 1.0\;{\rm cm}^{-3}$, the
dust mass ratios produced by AGB stars and SNe~II are 
$1.6 \times 10^{-3}$, $0.16$, and $0.37$ at $t = 0.1, 1.0$
and $10$~Gyr, respectively.
On the other hand, \citet{valiante09} suggested that the contribution
of AGB stars to the total dust mass in galaxies approaches or exceeds
that of SNe~II at $t \sim 1$~Gyr.
This difference mainly results from the dust mass data adopted.
We adopt the data of \citet{nozawa07}, whereas \citet{valiante09} adopted those of \citet{bianchi}.
The dust mass of \citet{nozawa07} is larger than that of
\citet{bianchi} because of the difference in the
treatment of the dust condensation and the destruction by reverse shocks.
However, even if the contribution of AGB stars is larger, 
the result that only a small amount of grains with $\la 0.01\;\mu$m are produced by stars does not change.

From the right panel in Fig.~\ref{fig:star_chansn}, we find that the
contribution of dust from AGB stars is relatively large for $n_{\rm SN} =
10.0\;{\rm cm}^{-3}$ at $t = 10$~Gyr.
At $t = 10$~Gyr, the dust mass ratios produced by AGB stars and SNe~II are $0.16$, $0.37$, and $1.39$ for the cases with $n_{\rm SN} =
0.1, 1.0$ and $10.0\;{\rm cm}^{-3}$, respectively.
This is because a larger amount of dust grains condensed in the
ejecta of SNe~II
are destroyed by reverse shocks for higher $n_{\rm SN}$.

\subsubsection{Dust destruction by SN shocks in the ISM}

\begin{figure}
\centering\includegraphics[width=0.45\textwidth]{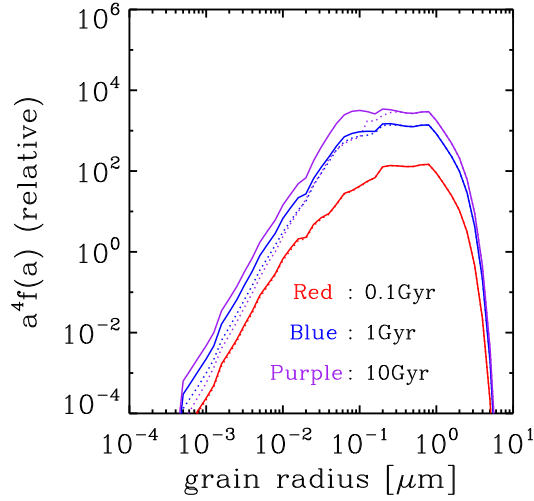}
\caption{Grain size distribution taking into account the dust destruction by SN shocks in the ISM in addition
 to the processes in Fig.~\ref{fig:staronly} (dotted lines). The values of $\tau_{\rm SF}$ and $n_{\rm SN}$ are the same as in
 Fig.~\ref{fig:staronly}.
Solid lines are the same as in Fig.~\ref{fig:staronly},
 the same color corresponding to the same age. Note that the red dotted
 line overlaps with the red solid line.}
\label{fig:sndest}
\end{figure}
In Fig.~\ref{fig:sndest}, we show the evolution of the grain size
distribution taking into account dust destruction by SN shocks in the
ISM in addition to the dust production by SNe~II and AGB stars.
We also present the cases without the dust destruction (i.e., the same as the
solid lines of Fig.~\ref{fig:staronly}).
The values of $\tau_{\rm SF}$ and $n_{\rm SN}$ are
set to the same values as in Fig.~\ref{fig:staronly}.
At $t \la 1.0$~Gyr, the grain size distributions with and without the
dust destruction by SN shocks in the ISM are very similar to each other.

Now we estimate the dust destruction timescale. First, we introduce the sweeping timescale, $\tau_{\rm sweep}$, at which SN
shocks sweep the whole ISM, as
\begin{equation}
\tau_{\rm sweep} \equiv \frac{M_{\rm ISM}}{M_{\rm swept} \gamma_{\rm SN}}.
\end{equation}
From Eq.~(\ref{eq:snrate}), if $C$ is defined as
\begin{equation}
C \equiv \int^{40\;{\rm M}_{\odot}}_{8\;{\rm M}_{\odot}} \phi(m) {\rm d}m,
\end{equation}
Eq.~(\ref{eq:snrate}), with Eq.~(\ref{eq:sfr}), can be approximated as
\begin{equation}
\gamma_{\rm SN} \simeq C \frac{M_{\rm ISM}}{\tau_{\rm SF}},
\end{equation}
where $C$ is about $1.5 \times 10^{-2}$ from our calculation.
Thus, if $n_{\rm SN} = 1.0\;{\rm cm}^{-3}$, $\tau_{\rm sweep} \sim
2${--}$4 \times 10^{-2} \tau_{\rm SF}$.
Next, we approximate the dust destruction rate by introducing the dust
destruction timescale, $\tau_{\rm SN}$, as 
\begin{equation}
\left.\frac{{\rm d}M_{\rm d}}{{\rm d}t}\right|_{\rm SN} \sim -\frac{M_{\rm
d}}{\tau_{\rm SN}}.
\label{eq:sndestcontri}
\end{equation}
The right hand side of Eq.~(\ref{eq:sndest}) can be approximated
as $- \tau^{-1}_{\rm sweep} M_{\rm d} (1 - \xi)$ where $\xi$ is a
typical value of $\xi_{\rm X}(a,a^{'})$;
then, Eq.~(\ref{eq:sndest}) reduces to
\begin{equation}
\tau_{\rm SN} \sim (1 - \xi)^{-1} \tau_{\rm sweep}.
\end{equation}
Since the overall efficiency of dust destruction, $(1 - \xi)$, is $\sim 0.3$ for $n_{\rm
SN} = 1.0\;{\rm cm}^{-3}$ \citep{nozawa06}, we obtain $\tau_{\rm SN} \sim 0.1 \tau_{\rm SF}$.
Thus, the difference between the cases with and without dust destruction cannot be seen at $t = 0.1$~Gyr
in Fig.~\ref{fig:sndest}, where $\tau_{\rm SN} \sim 0.1 \tau_{\rm SF}
\sim 0.5$~Gyr.

We find that dust grains with $a \la 0.1~\mu$m are destroyed
effectively at $10$~Gyr $\gg \tau_{\rm SN}$ (compare the solid and dotted lines in
Fig.~\ref{fig:sndest}).
Since the decreasing rate of grain radius by sputtering does not depends on the grain radius,
smaller grains are effectively destroyed in SN shocks, and the amount
of smaller grains decreases \citep{nozawa06}.

\begin{figure*}
\centering\includegraphics[width=0.45\textwidth]{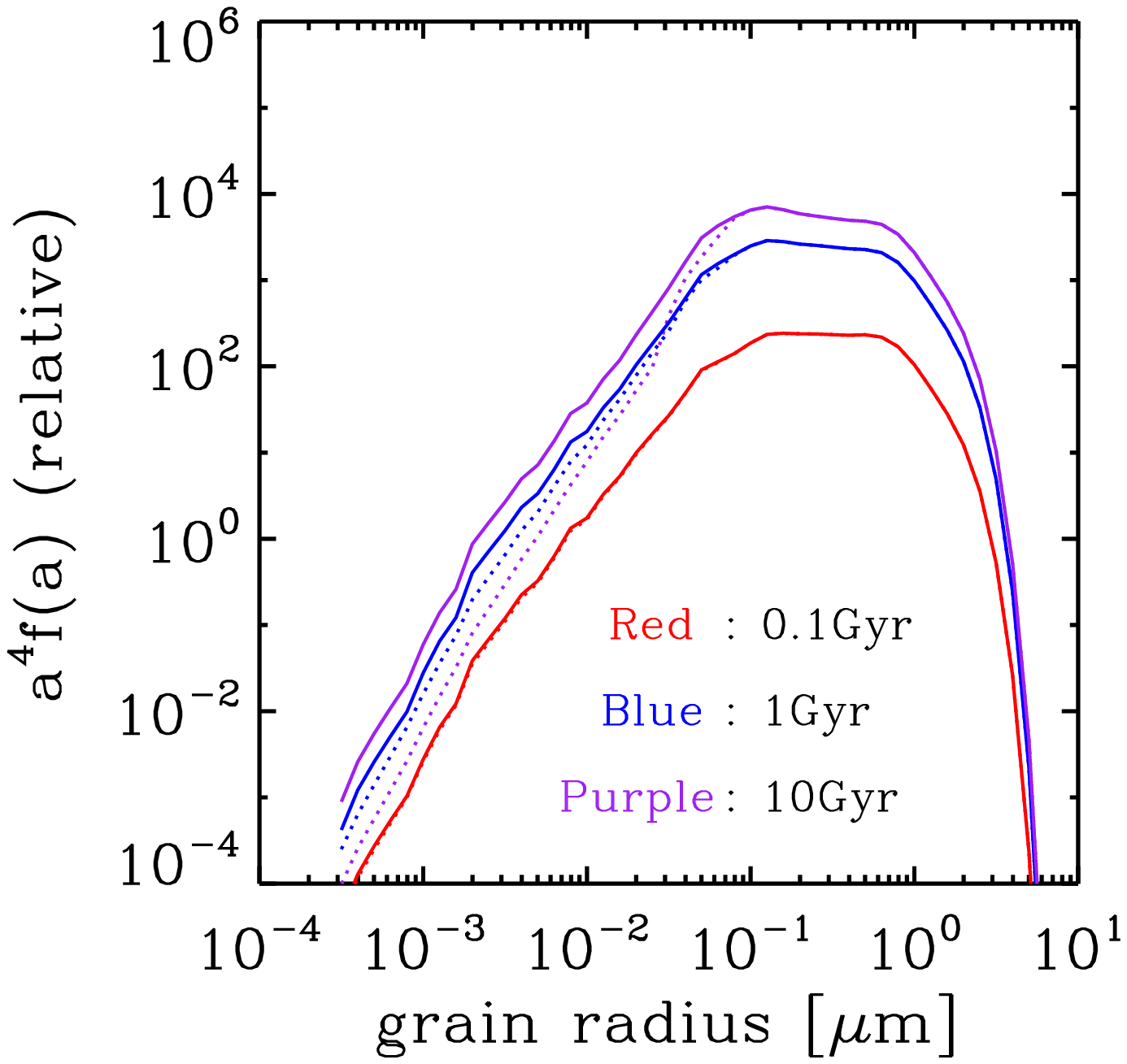}
\includegraphics[width=0.45\textwidth]{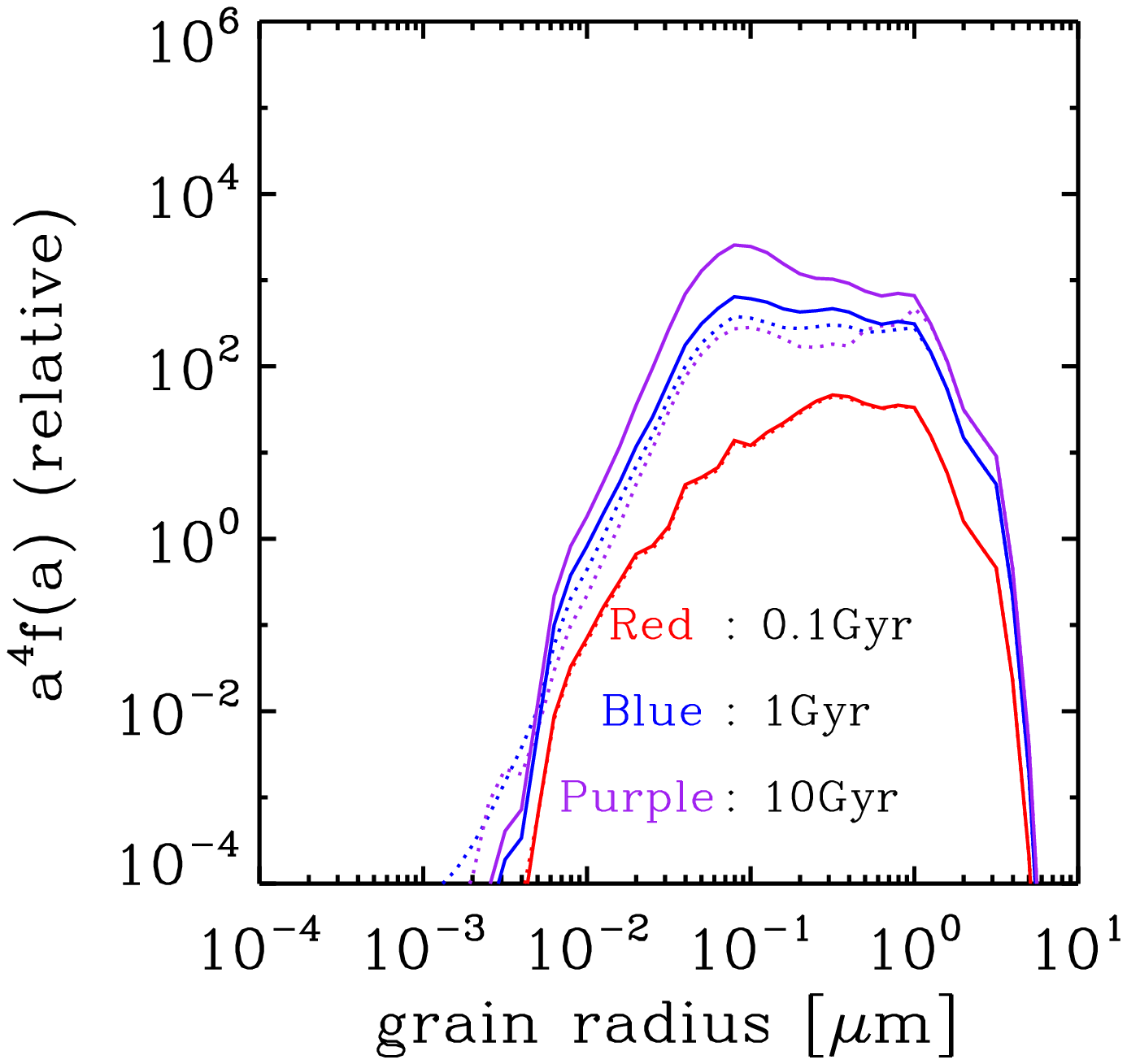}
\caption{Same as in Fig.~\ref{fig:sndest}, but we adopt different values
 of $n_{\rm SN}$: $0.1~{\rm cm}^{-3}$ in the left panel and $10.0~{\rm
 cm}^{-3}$ in the right panel.
Note that the red dotted lines overlap with the red solid lines.}
\label{fig:channsn}
\end{figure*}
In Fig.~\ref{fig:channsn}, we show the cases with $n_{\rm SN} = 0.1$ and
$10\;{\rm cm}^{-3}$.
Comparing figs.~\ref{fig:sndest} and \ref{fig:channsn}, we find that a
larger amount of dust grains are destroyed for higher $n_{\rm SN}$.
The destruction effect is more pronounced at small sizes.
Indeed, we observe that grains with $a \la 1.0~\mu$m are effectively
destroyed in the case with $n_{\rm SN} = 10.0~\mbox{cm}^{-3}$.
Nevertheless, the result that smaller grains are effectively
destroyed does not change, and we find that dust grains with radii larger than $0.1~\mu$m mainly dominate the total dust amount in galaxies.
Consequently, if the dust destruction by
sputtering in SN shocks is dominant, only large ($a \ga 0.1\;\mu$m) grains 
can survive in the ISM.

\subsubsection{Grain growth}

\begin{figure}
\centering\includegraphics[width=0.45\textwidth]{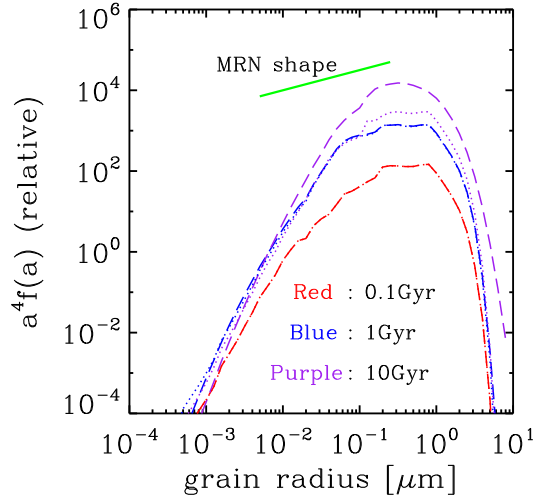}
\caption{Grain size distribution taking into account the dust production
 from stellar sources, dust destruction, and grain growth with
 $\tau_{\rm SF} = 5$~Gyr and $n_{\rm SN} = 1.0\;{\rm cm}^{-3}$
 (long-dashed lines).
We adopt $\eta_{\rm CNM} = 0.5$.
Dotted lines represent cases without grain growth [the same as in Fig.~(\ref{fig:sndest})], the same color
 corresponding to the same age. Note that the red dotted line overlaps
 with the red long-dashed line.
Green solid line represents the slope of the power-law grain size
 distribution with index $-3.5$ [$f(a){\rm d}a \propto a^{-3.5}{\rm d}a$ \citep{MRN}] which
 is thought to be the grain size distribution in the Milky Way.
}
\label{fig:eta0.5}
\end{figure}

Figure~\ref{fig:eta0.5} shows the evolution of the grain size
distribution taking into account the dust production from
stellar sources, dust destruction, and grain growth.
We adopt $\tau_{\rm SF} = 5$~Gyr, $n_{\rm SN} = 1.0~{\rm cm}^{-3}$, and the mass
fraction for the CNM, $\eta_{\rm CNM} = 0.5$.
From Fig.~\ref{fig:eta0.5}, we observe that while the grain size distributions with and
without grain growth are almost the same at ages $0.1$~Gyr and $1$~Gyr,
the difference is clear at $10$~Gyr.
The effect of grain growth is prominent around $a \sim 0.3~\mu$m at
$10$~Gyr, since the total surface area of grains is dominated by grains
with $a \sim 0.3\;\mu$m.
The timescale of grain growth is discussed in detail in
Section \ref{sec:discussion}.

In Fig.~\ref{fig:eta0.5}, we also plot the slope of the MRN size distribution, $f(a){\rm d}a \propto
a^{-3.5}{\rm d}a$ \citep{MRN}, which is thought to be the grain size distribution in
the Milky Way.
From Fig.~\ref{fig:eta0.5}, it is clear that the small grains with $a \la
0.01\;\mu$m are too few to reproduce the MRN size distribution.
However, the existence of the $70~\mu$m excess is considered to be a proof of the
existence of small grains \citep{bernard}.
Furthermore, \citet{takeuchi03,takeuchi05} argued by using their
infrared SED model that small grains are necessary to reproduce the near{--}mid
infrared SEDs of star forming galaxies.
Consequently, when we consider the case in which dust production by
SNe~II and AGB stars, dust
destruction, and grain growth take place, the grain size
distribution is always dominated by large grains, and we need to
consider other processes to produce small grains efficiently.

\subsection{Grain{--}grain collision effects}
\label{subsec:w/ collision}

In the above we have investigated the dust processes
which directly affect the evolution of the total
dust mass in galaxies: dust production by AGB stars and SNe~II, dust
destruction by SN shocks, and grain growth.
As shown above, these processes cannot produce
small grains $(a \la 0.01\;\mu{\rm m})$ efficiently.
Therefore, we now consider the contributions of the grain{--}grain collisions, shattering and
coagulation in turbulence, to the grain size distribution.
If these processes occur, although the total dust mass in galaxies
does not change, the grain size distribution does.

\subsubsection{Shattering}
\label{subsec:resultshat}

\begin{figure*}
\centering\includegraphics[width=0.45\textwidth]{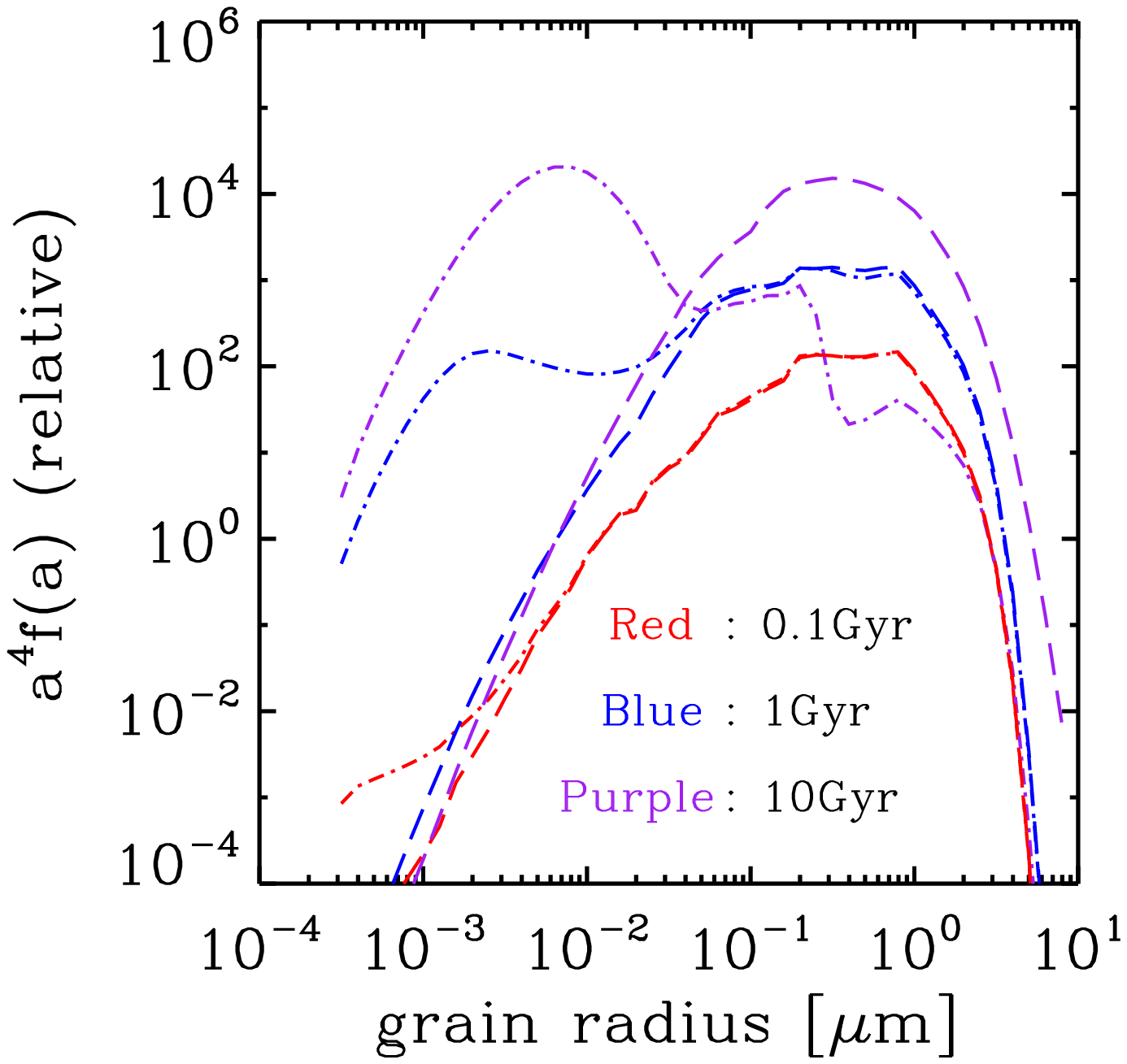}
\includegraphics[width=0.45\textwidth]{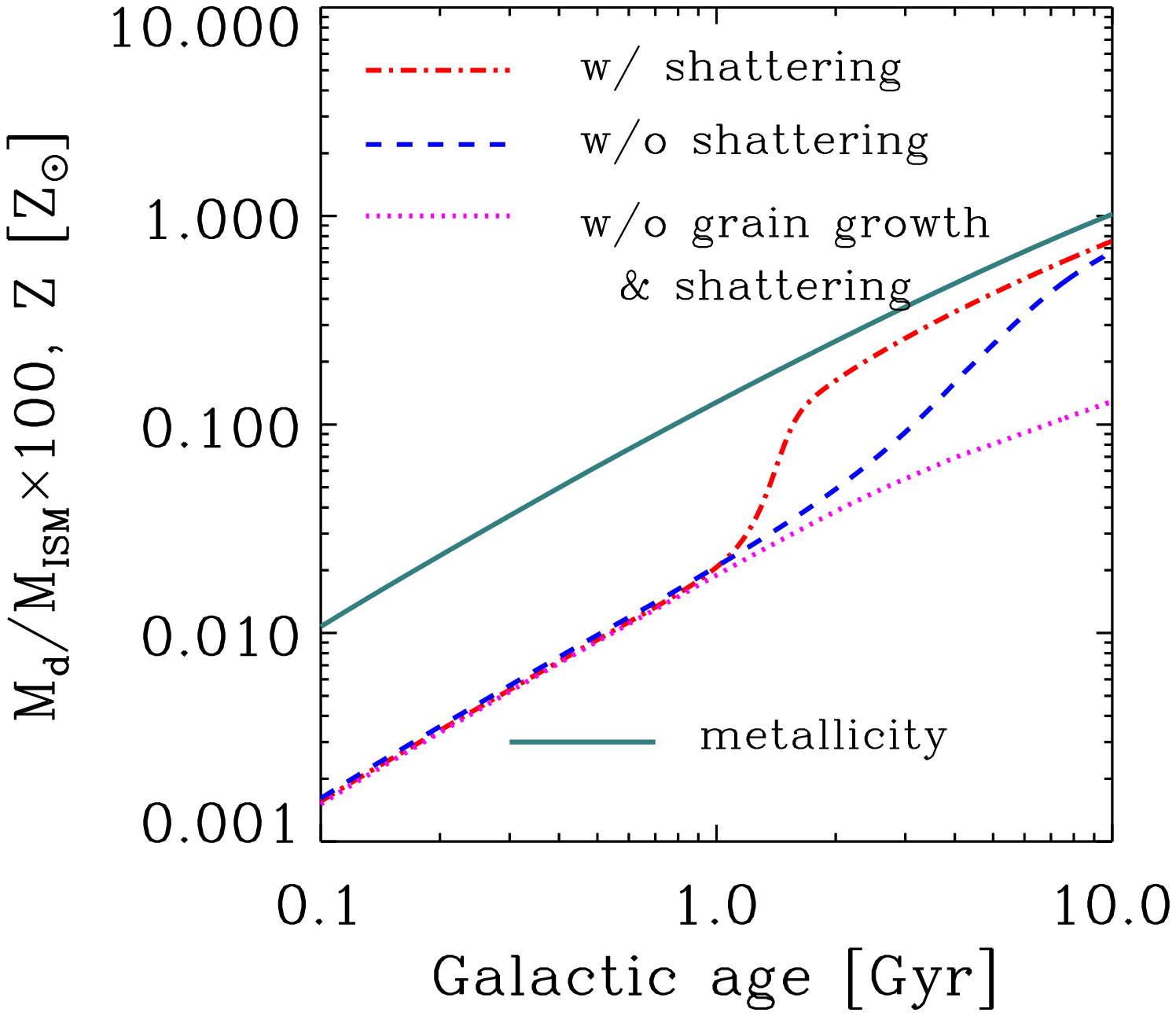}
\caption{Left panel: grain size distribution with (dot-dashed lines) and
 without (long-dashed lines) shattering (all other dust processes in
 Section~\ref{subsec:w/o collision} are included).
Note that the red dot-dashed line overlaps with the red long-dashed line. 
Right panel: time evolution of dust-to-gas mass ratio
 with (solid line) and without (dotted line) shattering.
The case without grain growth and shattering (dashed line) is also
 plotted. Dot-dashed line represents the evolution of metallicity.
The parameters $\tau_{\rm SF}$ and $n_{\rm SN}$ are set to be
 $5~\mbox{Gyr}$ and $1.0~{\rm cm}^{-3}$, respectively.
We adopt $\eta_{\rm WNM} = \eta_{\rm CNM} = 0.5$.
}
\label{fig:shattering}
\end{figure*}
In the left panel of Fig.~\ref{fig:shattering}, we show the evolution of the grain size
distribution in the galaxy with and without shattering (all other dust
processes in Section~\ref{subsec:w/o collision} are included).
The right panel of Fig.~\ref{fig:shattering} shows the time evolution of
dust-to-gas mass ratio ($M_{\rm d}/M_{\rm ISM}$) for the cases with and without shattering, respectively.
We also plot the case without grain growth and shattering and the
evolution of metallicity in the same panel.
The parameters $\tau_{\rm SF}$ and $n_{\rm SN}$ are set
to be $5$~Gyr and $1.0\;{\rm cm}^{-3}$, respectively.
We adopt $\eta_{\rm WNM} = \eta_{\rm CNM} = 0.5$.

From the left panel of Fig.~\ref{fig:shattering}, 
at the early stage of galaxy evolution ($0.1$~Gyr) the size
distributions with and without shattering are similar with only a little
difference at small sizes.
At $1$~Gyr, we observe that the size distribution has a bump at $a \sim 0.001~\mu$m in the case with shattering.
As time passes, the amount of large grains decreases, and as a result the size
distribution is dominated by small grains.
This behavior is substantially different from that of the case without shattering.
We now discuss this behavior in more detail.

As shown in Eq.~(\ref{eq:shattering}), the efficiency of shattering is
larger for larger amount of grains \citep{hirashita10a}.
At $0.1$~Gyr, the efficiency of shattering is low because of the small
dust abundance.
As a result, there is only a small difference between the cases with and
without shattering.
At $t = 1$~Gyr, since shattering occurs efficiently due to the increased
amount of large grains, the amount of small grains increases.
At the same time, we observe that the grain size distribution has little difference between the cases with and
without shattering at $a > 0.1~\mu$m in the left panel of Fig.~\ref{fig:shattering}.
This is because shattering of a tiny fraction of large grains can
produce a large amount of small grains \citep{hirashita10b}.
Furthermore, since the number of small grains increases,
the small grains dominate the total grain surface area.
Consequently, grain growth occurs at the smallest grain sizes ($a \la
10^{-3}\;\mu$m), forming a bump at $\sim 10^{-3}$--$10^{-2}\;\mu$m.
At $t = 10$~Gyr, since the number of small grains increases,
large grains are shattered more efficiently by the frequent collisions
with the small grains.
Consequently, comparing the grain size distribution at $10$~Gyr with that at $1$~Gyr, the amount of large
grains decreases significantly.
Furthermore, because of grain growth, the bump is shifted to a larger
size at $10$~Gyr than at $1$~Gyr, and finally the size distribution has
a large bump at $a \sim 0.01~\mu$m at $10$~Gyr.

%discussion about maximum size of the size distribution
Focusing on the grain size distribution at $10$~Gyr, we find that if shattering occurs, the amount of grains with $a >
0.2\;\mu$m is more than two orders of magnitude smaller than that of
grains with $a < 0.2\;\mu$m.
Thus, the maximum size of grains in diffuse ISM is determined not by
stardust but by the process of shattering.

In the right panel of Fig.~\ref{fig:shattering}, we find that grain
growth starts to increase the total dust mass at around $t = 1$~Gyr as seen in the
rapid increase of dust-to-gas mass ratio,
 and grain growth becomes more rapid in the case with shattering than in
 the case without shattering because of the increased number of small grains.
As discussed in \citet{kuo}, shattering contributes not only to the
evolution of the grain size distribution but also to the total dust
mass in galaxies indirectly through the enhanced grain growth.
Thus, shattering is a very important process in understanding the evolution of the size distribution and the amount of dust
grains in the ISM.

\subsubsection{Coagulation}
\label{subsubsec:resultcoag}

\begin{figure*}
\centering\includegraphics[width=0.45\textwidth]{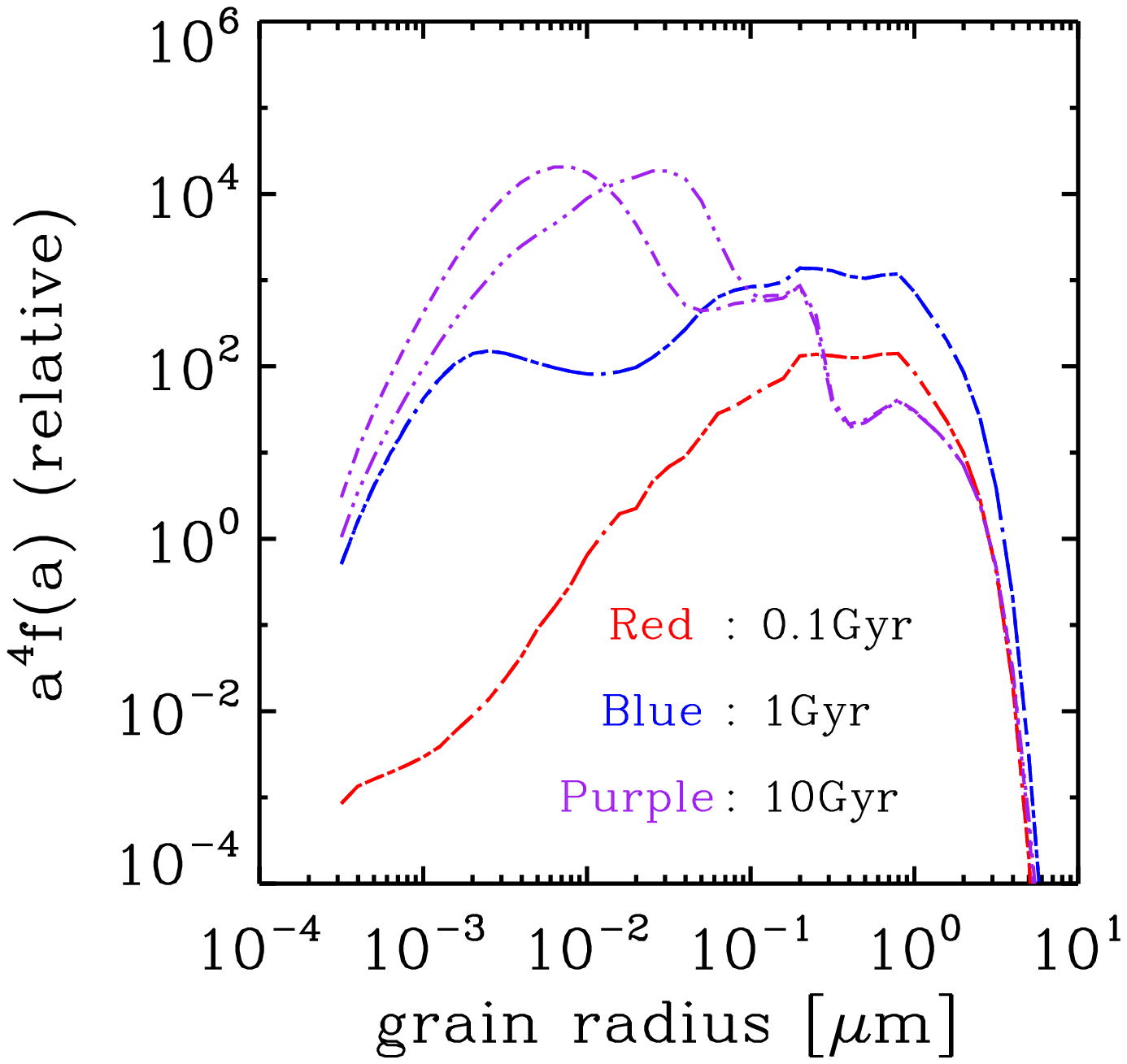}
\includegraphics[width=0.45\textwidth]{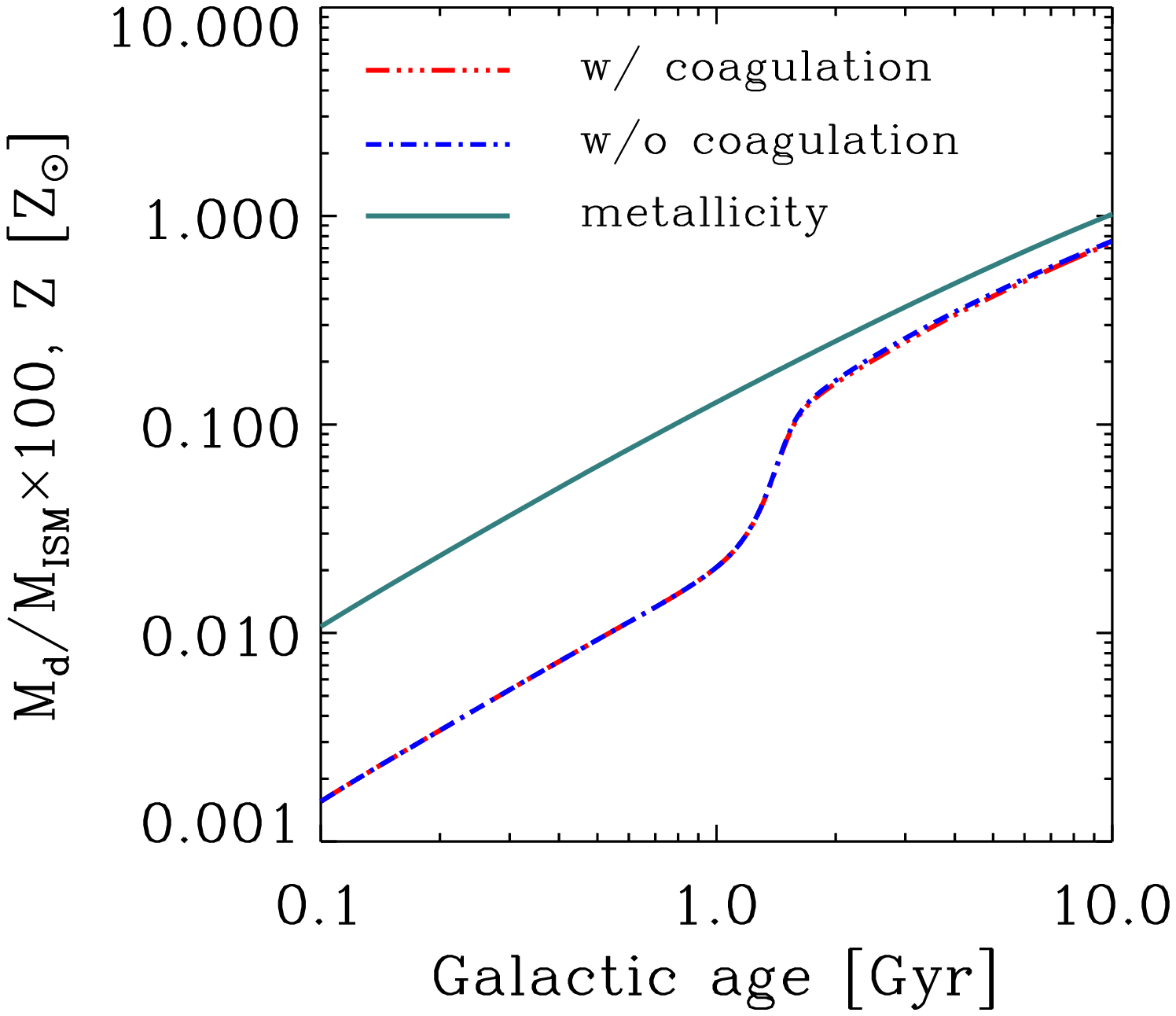}
\caption{Left panel: grain size distribution with (triple-dot-dashed lines) and
 without (dot-dashed lines) coagulation (all the other dust processes are included).
Note that the red and blue dot-dashed lines overlap with the red and blue triple-dot-dashed lines.
Right panel: time evolution of dust-to-gas mass ratio with (solid line)
 and without (dotted line) coagulation.
Dot-dashed line represents the evolution of metallicity.
The values of parameters ($\tau_{\rm SF}, n_{\rm SN},
 \eta_{\rm WNM}$, and $\eta_{\rm CNM}$) are the same as in Fig.~\ref{fig:shattering}.
}
\label{fig:coagulation}
\end{figure*}
In Fig.~\ref{fig:coagulation}, we show the evolution
of the grain size distribution with and without coagulation (all the
other dust processes are included) in the left panel,
and the time evolutions of dust-to-gas mass ratio with and without
coagulation and of metallicity in the right panel.
The parameters adopted are the same as in
Fig.~\ref{fig:shattering}.

From the left panel of Fig.~\ref{fig:coagulation},
we find that there is little difference between the cases with and
without coagulation at $0.1$ and $1.0$~Gyr.
Since larger grains are coupled with the larger-scale turbulence, they can obtain larger velocity dispersions.
Thus, coagulation mainly occurs by collisions between small grains whose
velocity dispersions are smaller than the coagulation threshold [Eq.~(\ref{eq:coagthrevel})].
However, since the abundance of small grains is low, the contribution of
coagulation is not seen at $0.1$ and $1$~Gyr before shattering becomes effective.
After that, a large abundance of small grains are produced by shattering so coagulation becomes effective.
Consequently, the bump at $a \sim 0.01~\mu$m shifts to a larger size by coagulation.

From the right panel of Fig.~\ref{fig:coagulation}, we find that the evolution of the total dust mass does not change
significantly by coagulation, confirming the result obtained by \citet{hirashita12a}.
If coagulation occurs, the number of small grains decreases; as a
result, the surface-to-volume ratio of grains decreases.
This effect may suppress the increase in dust mass due to grain growth.
However, since grain growth becomes inefficient to the dust evolution
before coagulation becomes efficient (the details are shown in
Section \ref{sec:discussion}),
the contribution of coagulation cannot be observed for the total dust
mass evolution. 
Consequently, the effect of coagulation on the evolution of the total dust mass in galaxies is negligible.

{}From the left panel of Fig.~\ref{fig:coagulation}, 
we find that the amount of grains with $a > 0.2~\mu$m does not change
significantly by coagulation because coagulation cannot occur by collision between large grains which
have larger velocity dispersions than the coagulation threshold.
Thus, although the bump of the grain size distribution is shifted to a
larger size by coagulation, coagulation does not affect the maximum size determined by
shattering (Section~\ref{subsec:resultshat}).

\subsection{Parameter dependence}
\label{subsec:paradep}

Shattering and coagulation occur differently in both ISM phases (WNM and CNM).
Here, by adopting $(\eta_{\rm WNM}, \eta_{\rm CNM}) = (0.9, 0.1)$ and
$(0.1, 0.9)$, we show the effect of ISM phases.

\begin{figure*}
\centering\includegraphics[width=0.45\textwidth]{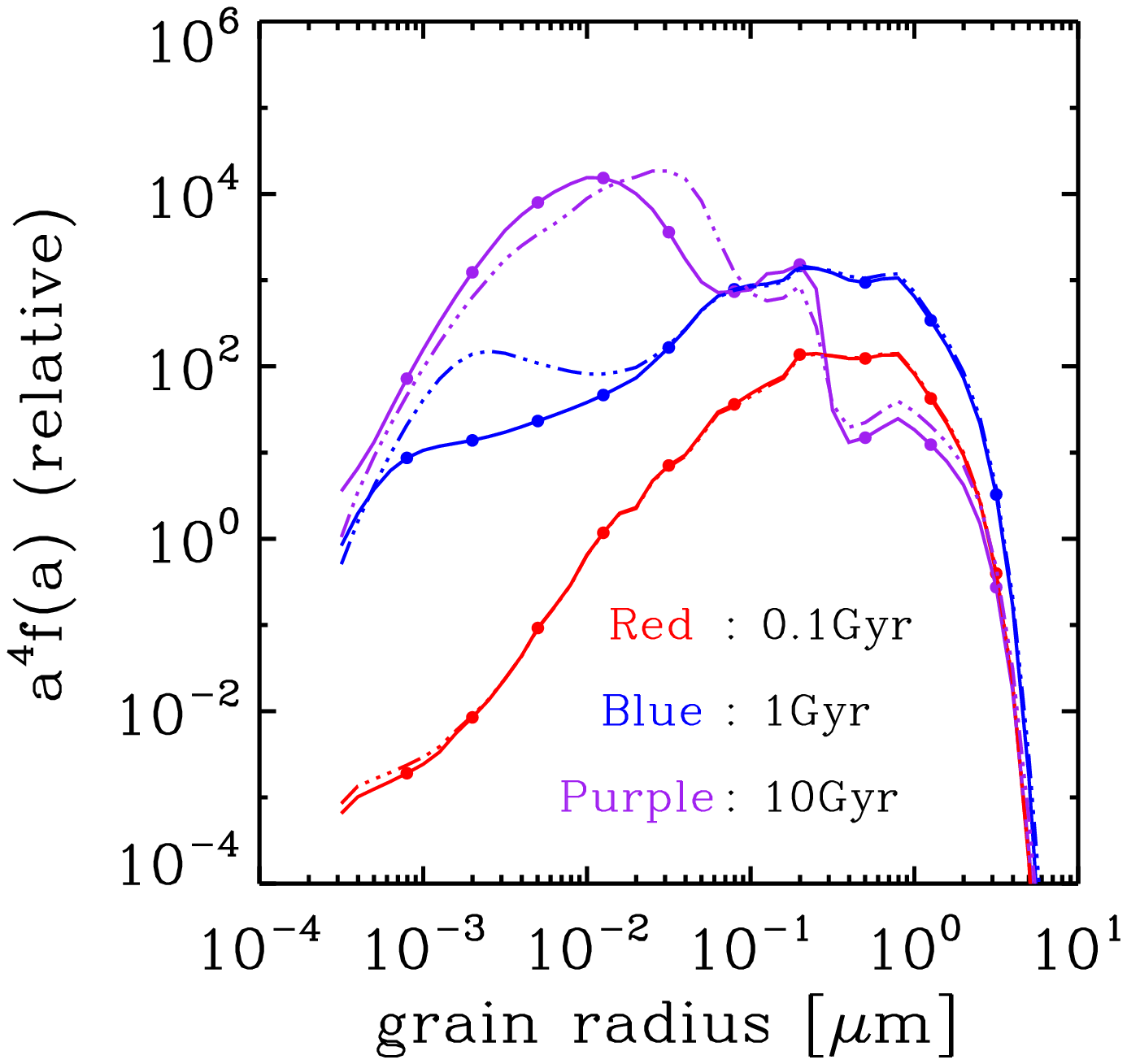}
\includegraphics[width=0.45\textwidth]{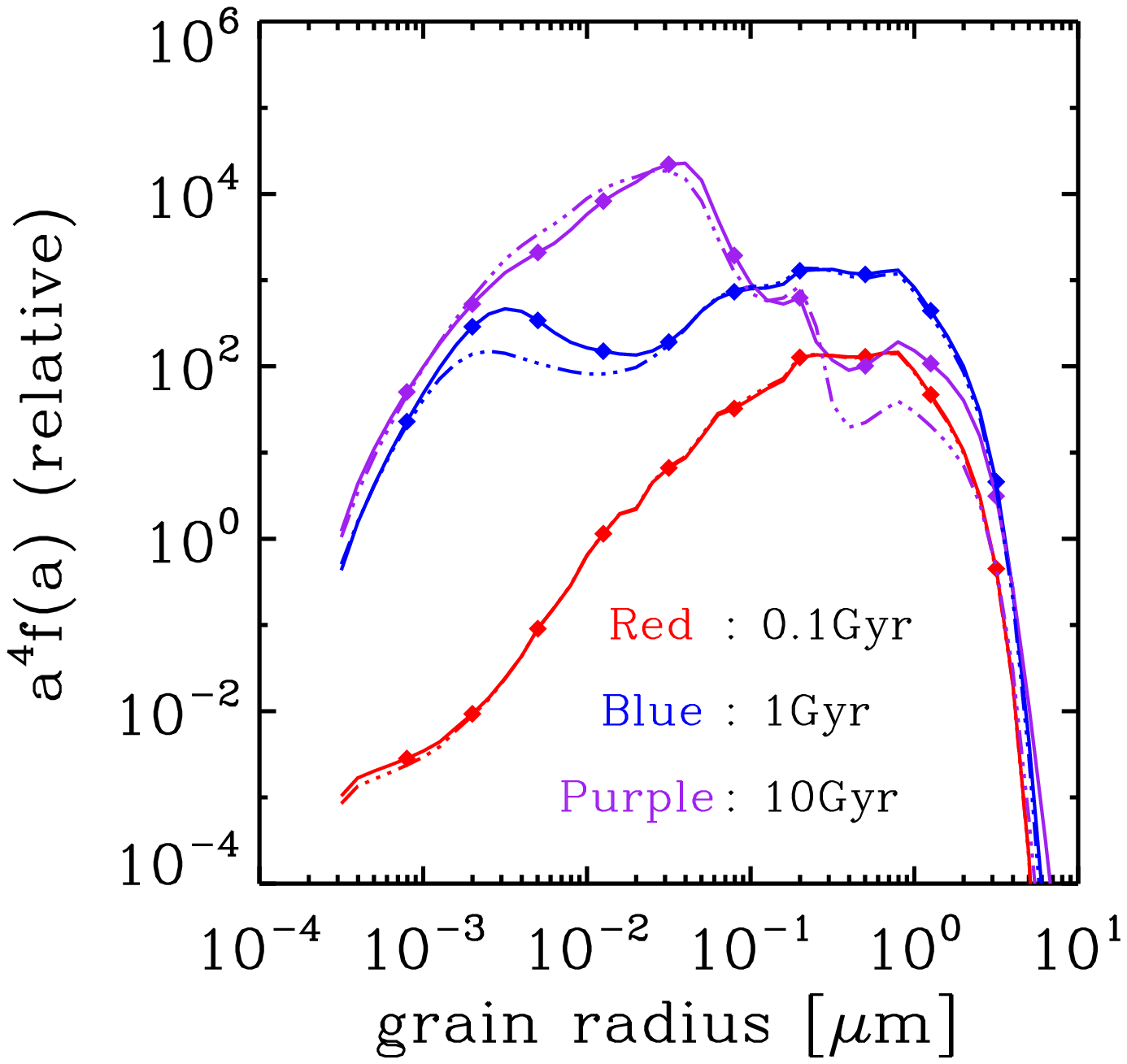}
\caption{Grain size distribution with $(\eta_{\rm WNM}, \eta_{\rm CNM})
 = (0.9, 0.1)$ (solid lines with filled circles in the left panel) and
 $(0.1, 0.9)$ (solid lines with filled diamonds in the right panel).
The values of parameters $\tau_{\rm SF}$ and $n_{\rm SN}$ are
 the same as in Fig.~\ref{fig:coagulation}.
Triple-dot-dashed lines are the case with $\eta_{\rm WNM} = \eta_{\rm CNM} = 0.5$.
}
\label{fig:paradep}
\end{figure*}
In Fig.~\ref{fig:paradep}, we show the evolution of the grain size
distribution with $(\eta_{\rm WNM}, \eta_{\rm CNM}) = (0.9, 0.1)$ (left
panel) and $(0.1, 0.9)$ (right panel).
The case with $\eta_{\rm WNM} = \eta_{\rm CNM} =
0.5$ is shown for comparison.
At $t = 0.1$~Gyr, the grain size distributions are almost the same in
all the cases, since the dust process is dominated by the production by
stellar sources.
At $t = 1$~Gyr, the amount of dust grains with $a < 0.01\;\mu$m is larger for the case with larger
$\eta_{\rm CNM}$ because grain growth is more efficient.
At $10$~Gyr, the difference is clear at each grain size.
For larger $\eta_{\rm WNM}$, the decrements of the amount of dust grains
with $a > 0.2\;\mu$m is larger because shattering in the WNM is more efficient.
Furthermore, for larger $\eta_{\rm CNM}$, the bump produced by grain growth around $0.01\;\mu$m
shifts to a larger size. Thus, we understand that the amount of dust grains with $a >
0.2\;\mu$m and the shift of the bump around $0.01\;\mu$m are
dominated by shattering in WNM and coagulation in CNM, respectively.
In addition, comparing the two panels in Fig. 8, we
find that the dust amount at $a \sim 0.1$--$0.2\;\mu$m tends to
be smaller for a larger $\eta_{\rm CNM}$ \footnote{The mass
ratio of grains with $a \sim 0.1$--$0.2\;\mu$m for the cases between
($\eta_{\rm WNM}, \eta_{\rm CNM}$) = $(0.5, 0.5)$ and $(0.9, 0.1)$ is
about $0.6$ at $t = 10$~Gyr.}.
It means that the amount of dust grains with $a \sim 0.1\;\mu$m is
dominated not by shattering in WNM but shattering in CNM.
Hence, the grain size distribution in galaxies is
finally dominated by processes in WNM for large grains ($> 0.2\;\mu$m) and by processes in CNM
for small grains ($\sim 0.1\;\mu$m).  

\section{Discussion}
\label{sec:discussion}
In Section~\ref{sec:result}, we showed the evolution of the
grain size distribution in galaxies for a variety of mixture of dust processes.
We found that the grain size distribution is
dominated by large grains produced by stars (SNe~II and AGB stars) in the early stage of galaxy
evolution, but as the time passes the number of small grains increases
due to shattering, and the small grains grow to larger grains by grain growth. After that, the size distribution shifts to larger sizes due to coagulation.
Thus, we conclude that, while the grain size distribution in galaxies
is controlled by stellar processes in the early stage of galaxy evolution, 
the main driver to change the size distribution is replaced with the processes
in the ISM (shattering, coagulation, and grain growth) at the later stage of galaxy evolution.
These processes (shattering, coagulation, and grain growth) have
timescales dependent on the grain size distribution.
In this Section, by adopting representative grain radii, $0.001\;\mu$m,
$0.01\;\mu$m, $0.1\;\mu$m, and $1.0\;\mu$m, we discuss the evolution of
the grain size distribution more quantitatively.

\begin{figure}
\centering\includegraphics[width=0.45\textwidth]{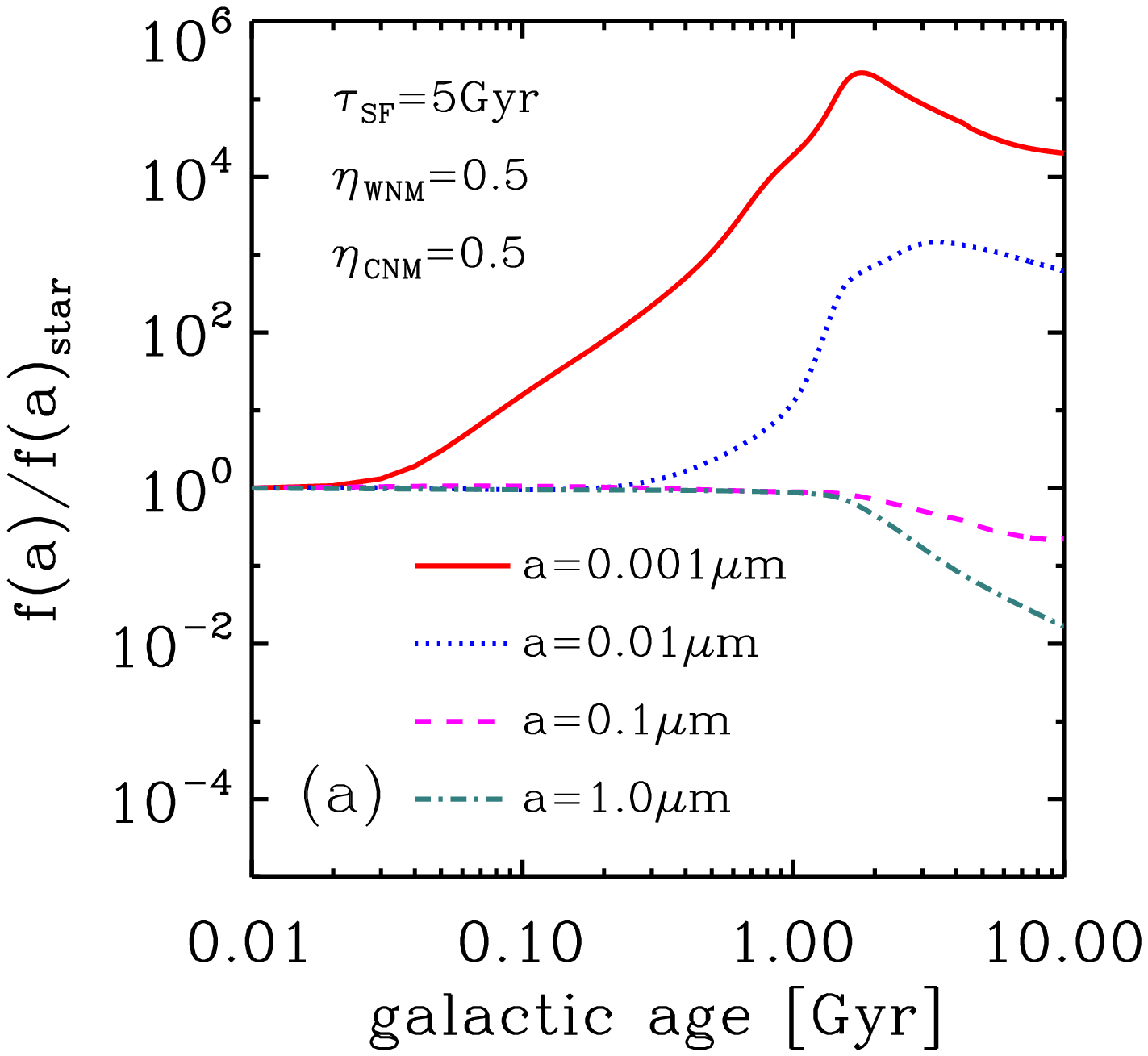}
\includegraphics[width=0.45\textwidth]{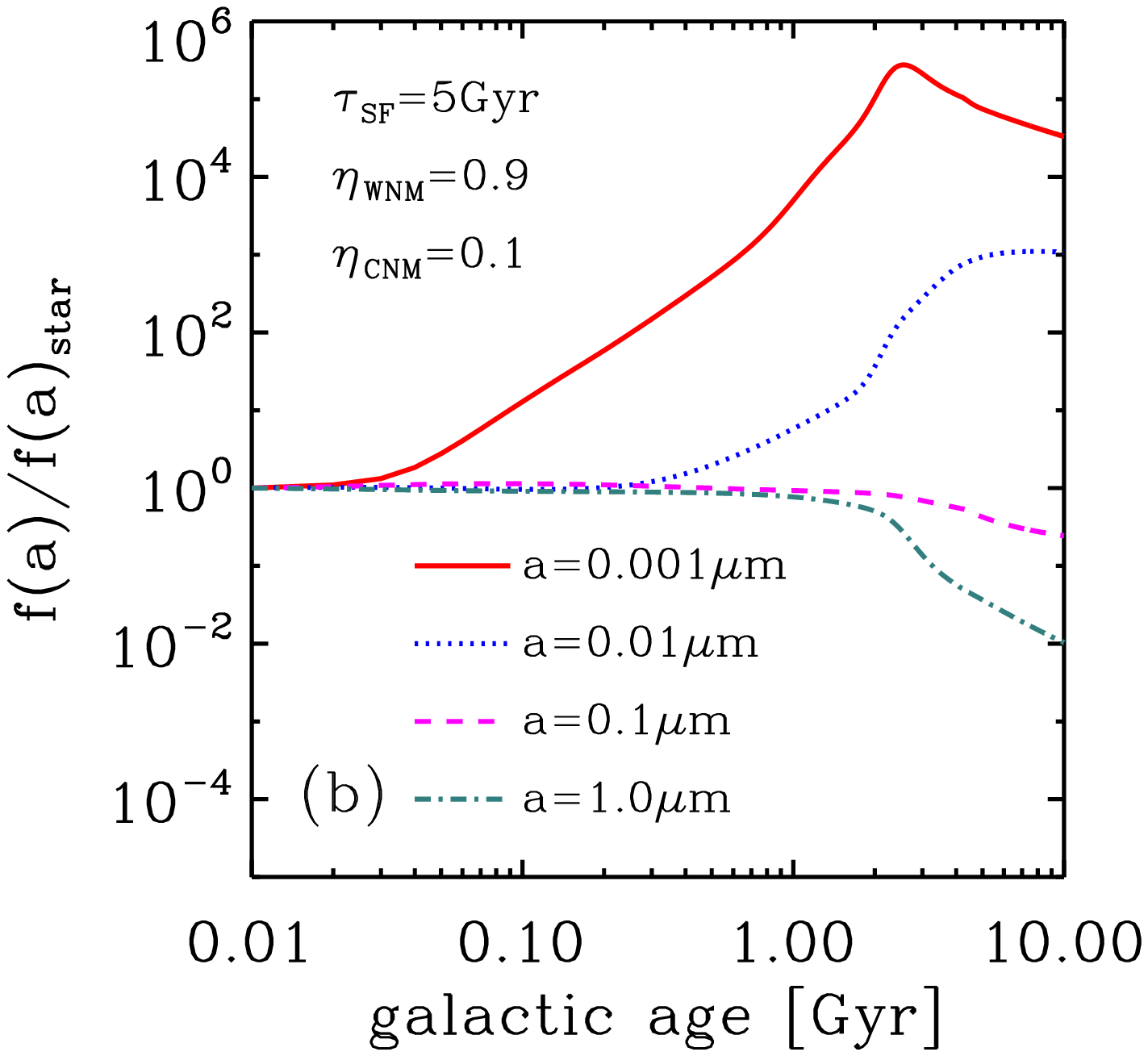}
\includegraphics[width=0.45\textwidth]{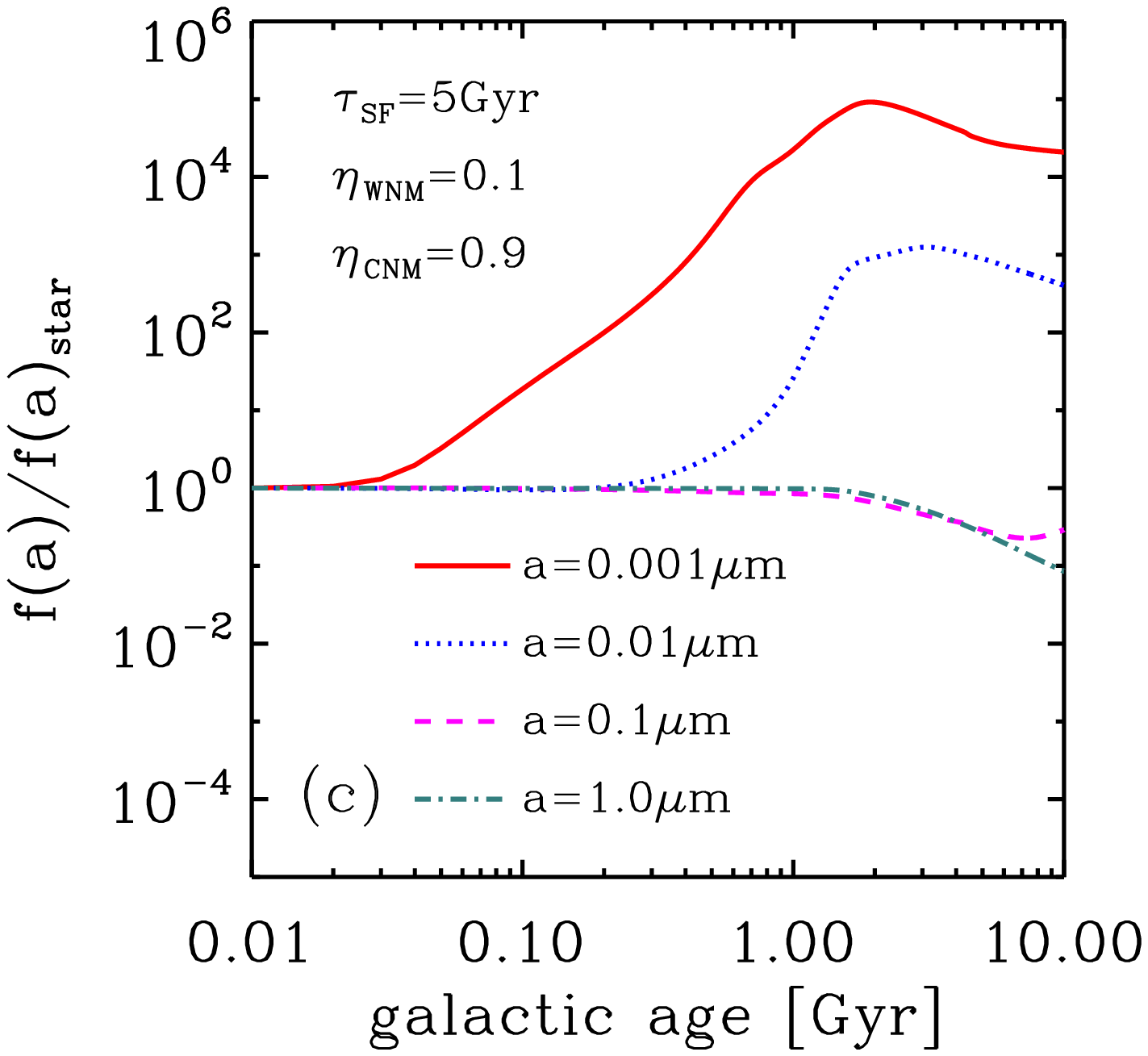}\\
\includegraphics[width=0.45\textwidth]{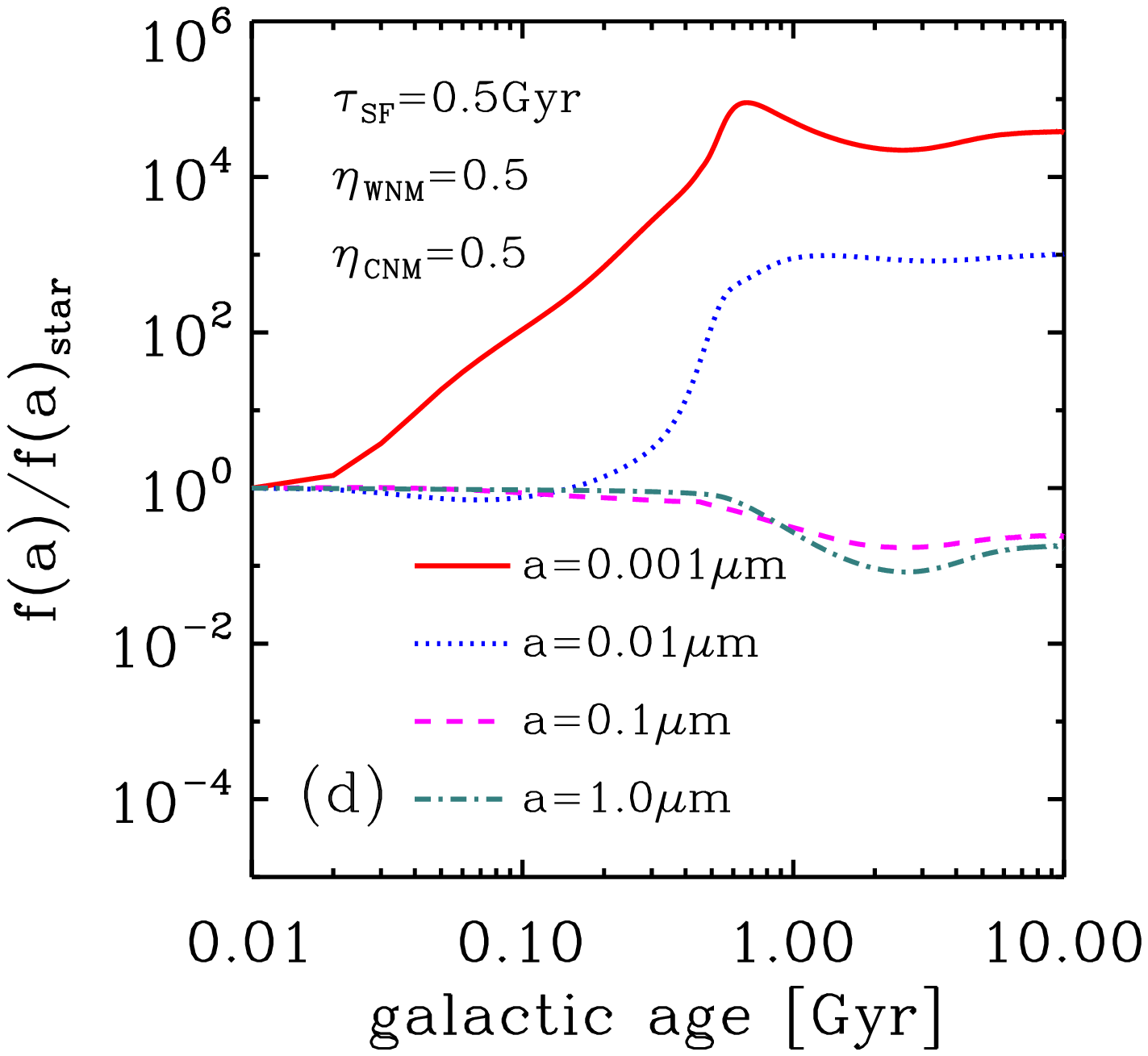}
\includegraphics[width=0.45\textwidth]{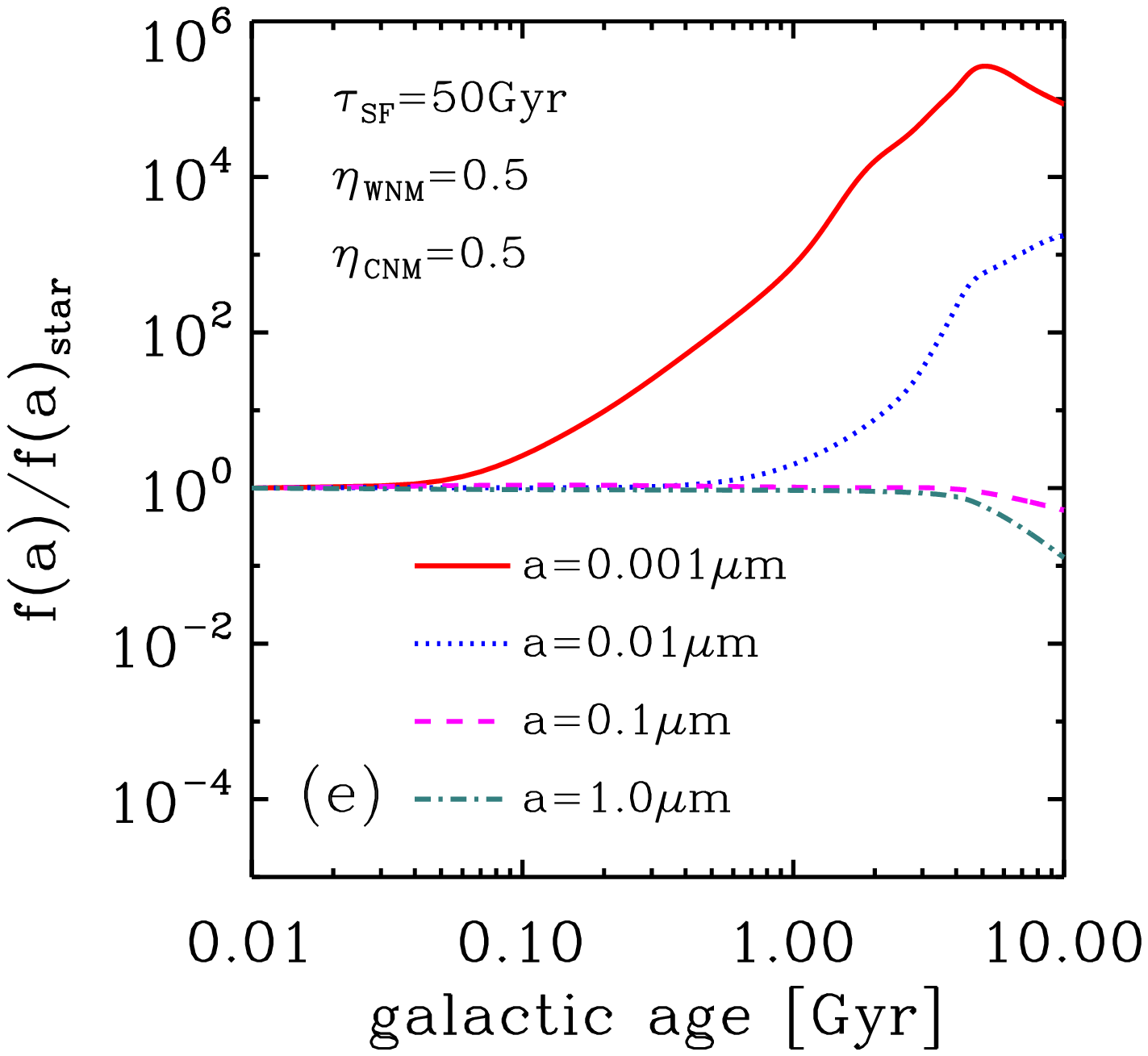}
\caption{Time evolution of the ratio between the size distribution
 functions, $f(a)$, and $f(a)_{\rm star}$, the latter being obtained by considering only stellar processes [the first and second terms in
 right hand size of Eq.~(\ref{eq:dustevo})]. 
Panel (a), (b), and (c) are the cases with $(\eta_{\rm WNM}, \eta_{\rm CNM}) = (0.5, 0.5), (0.9,
 0.1)$, and $(0.1, 0.9)$, respectively with $\tau_{\rm SF} = 5$~Gyr and $n_{\rm SN} = 1.0\;{\rm cm}^{-3}$.
Panel (d) and (e) are the case with $\tau_{\rm SF} = 0.5$~Gyr and
 $50$~Gyr, respectively with $(\eta_{\rm WNM}, \eta_{\rm CNM}) = (0.5, 0.5)$.
Solid, dotted, dashed, and dot-dashed lines represent the ratio of $a =
 0.001\;\mu$m, $0.01\;\mu$m, $0.1\;\mu$m, and $1.0\;\mu$m, respectively.
}
\label{fig:eachsizeevo}
\end{figure}

Figure~\ref{fig:eachsizeevo} shows the time evolution of the ratio
between the size distribution functions, $f(a)$, and $f(a)_{\rm star}$,
the latter being obtained by considering only the stellar processes [the first
and second terms in the right hand size of Eq.~(\ref{eq:dustevo})].
Panels (a), (b), and (c) are the cases with $(\eta_{\rm WNM}, \eta_{\rm
CNM}) = (0.5, 0.5), (0.9, 0.1)$, and $(0.1, 0.9)$, respectively, for $\tau_{\rm SF} =
5$~Gyr and $n_{\rm SN} = 1.0\;{\rm cm}^{-3}$.
From Fig.~\ref{fig:eachsizeevo}, we find that behavior of
$f(a)/f(a)_{\rm star}$ depends strongly on the grain radius.
As mentioned above, this is because each dust process works at different
grain radii on different timescales.
First, we discuss the evolution of the grain size in panel (a).
We find that $f(a)/f(a)_{\rm star}$ at $a = 0.001\;\mu$m starts to
deviate from unity at the earliest galactic age among all four grain sizes.
The process that causes this increase is shattering.
These small grains are produced by shattering between large grains
produced by stars.
Furthermore, as mentioned in Section \ref{subsec:resultshat}, since
shattering of a small number of large grains can produce a large number of small grains, we cannot see
the change of $f(a)/f(a)_{\rm star}$ for $a = 0.1$ and $1.0\;\mu$m.

At $t \sim$ a few hundreds of Myr, $f(a)/f(a)_{\rm star}$ at $a = 0.01\;\mu$m
increases.
This increase is also due to shattering.
The reason why the effect of shattering appears at $a =
0.01\;\mu$m later than at $a =
0.001\;\mu$m is that the size distribution of shattered fragments is
proportional to $a^{-3.3}$ (see Section \ref{subsec:shattering}).
In other words, the shattered fragments become dominant at smaller sizes
on shorter timescales than at larger sizes.

At $t \sim 1$~Gyr, we find that the increase of $f(a)/f(a)_{\rm star}$ at $a =
0.01\;\mu$m is accelerated.
This indicates that another process becomes efficient, and it is grain
growth. As seen from Fig.~\ref{fig:shattering}, as grain growth becomes
efficient around $1$~Gyr, the amount of grains with less than $a \sim
0.01\;\mu$m increases significantly.

At $t \sim 2$~Gyr, we find that $f(a)/f(a)_{\rm star}$ decrease at all sizes.
These decreases are due to coagulation for small grains
($a = 0.001$ and $0.01\;\mu$m) and shattering for large grains ($a = 0.1$ and
$1.0\;\mu$m).
As we showed in Section \ref{subsubsec:resultcoag}, coagulation mainly
occurs between small grains.
Thus, the coagulation effect cannot be seen at early phase of galaxy
evolution when the abundance of small ($a \la 0.01\;\mu$m) grains is small.
Shattering can also occur effectively if there is a large amount of small
grains because of a high grain{--}grain collision rate with small grains [cf.~Eq.~(\ref{eq:shatteredrate})].
In addition, the main reason why the decrements of
grains with $a = 0.1$ and $1.0\;\mu$m are different is shattering in
different ISM phases. 
As shown in Section \ref{subsec:paradep}, grains with $a >
0.2\;\mu$m are mainly dominated by shattering in WNM, while grains with
$a \sim 0.1\;\mu$m are dominated by shattering in CNM.
In summary, at early phase of galaxy evolution ($t \la 10$~Myr),
the size distribution is dominated by dust grains produced by stars,
after $t \ga 100$~Myr, the dust processes in the ISM begin to affect the size
distribution at small size, and at $t \sim 2$~Gyr (for $\tau_{\rm SF} =
5$~Gyr), various dust processes in the ISM affect all sizes of grains.

Panels (b) and (c) in Fig.~\ref{fig:eachsizeevo} show the cases with ($\eta_{\rm WNM}, \eta_{\rm
CNM}$) $= (0.9, 0.1)$ and (0.1, 0.9), respectively.
Compared with panel (a), we find that $f(a)/f(a)_{\rm star}$ at $a =
0.01\;\mu$m does not decrease at $10$~Gyr in panel (b).
This is because the timescale of coagulation becomes longer for
smaller $\eta_{\rm CNM}$.
From panel (c) ($\eta_{\rm WNM} = 0.1, \eta_{\rm CNM} = 0.9$), we find
that the decrement at $a = 1.0\;\mu$m is
smaller than those in the cases of panel (a) and (b).
This is because the efficiency of shattering in WNM is smaller for
smaller $\eta_{\rm WNM}$.
However, from all the three panels, we can observe that the timing at
which $f(a)/f(a)_{\rm star}$ at all sizes changes due to
the dust processes in the ISM (in this case, it is about $2$~Gyr) does
not vary significantly by the change of ($\eta_{\rm
WNM}, \eta_{\rm CNM}$) for the same star formation timescale.

In order to discuss the effect of $\tau_{\rm SF}$ on the size distribution,
the results are shown for the same values of
the parameters as in the panel (a) of Fig.~\ref{fig:eachsizeevo}, but for $\tau_{\rm
SF} = 0.5$~Gyr in panel (d) and $\tau_{\rm SF} = 50$~Gyr in panel (e).
Compared with panel (a), we find that $f(a)/f(a)_{\rm star}$
change at earlier stages for shorter $\tau_{\rm SF}$ at all sizes.
This is explained as follows.
If $\tau_{\rm SF}$ is short, the amounts of dust and metals released by
stars are large at early phases of galaxy evolution.
The timescales of shattering and coagulation are
inversely proportional to the dust-to-gas mass ratio \citep[e.g.,][]{hirashita10a,
hirashita09a}, and the timescale of grain growth is inversely proportional to
metallicity \citep[e.g.,][]{asano}.
Thus, for shorter $\tau_{\rm SF}$, dust processes in the ISM (grain growth, shattering, and
coagulation) begin to affect the size distribution at earlier stages
of galaxy evolution ($\sim 0.6$, $2$ and $5$~Gyr for $\tau_{\rm SF} =
0.5, 5$, and $50$~Gyr, respectively).
The timescale of the change of $f(a)/f(a)_{\rm star}$ is roughly
estimated to be $\sim 1$ $(\tau_{\rm SF}/\mbox{Gyr})^{1/2}$~Gyr (Appendix~\ref{app:timescale}).
We conclude that the grain size distribution in galaxies changes drastically through
the galaxy evolution because different dust processes operate on the grain
size distribution at different ages.

\section{Conclusions}
\label{sec:conclusion}

We constructed a dust evolution model taking into account the grain
size distribution in a galaxy, and investigated what kind of dust
processes dominate the grain size distribution at each stage of galaxy evolution.
In this paper, we considered dust formation by SNe~II and AGB stars,
dust destruction by SN shocks in the ISM, grain growth in the CNM, and
grain{--}grain collisions (shattering and coagulation) in the WNM and CNM.

We found that the grain size distribution in galaxies is dominated
by large grains produced by stars in the early stage of galaxy
evolution, but as time passes the size distribution is controlled by processes in the ISM (grain growth, shattering, and coagulation) and
the age at which these ISM processes enter depends on the star
formation timescale, as $\sim 1 (\tau_{\rm SF}/\mbox{Gyr})^{1/2}$~Gyr.
While dust production by SNe~II and AGB stars, dust destruction by SN shocks, and
grain growth in the CNM directly affect the total dust mass evolution,
we found that the grains are predominantly large ($a \sim
0.2$--$0.5~\mu$m) and only a small amount of small
grains ($a < 0.01~\mu$m) are produced by these processes.
If we take shattering and coagulation into account, the grain size
distribution is modified significantly by these two processes.
In particular, shattering indirectly contribute to the large increase of
the total dust mass: After small grains ($a \la 0.01\;\mu$m) are
produced by shattering, grain growth becomes more
effective because of the enhanced surface-to-volume ratio.
Furthermore, grain growth produces a large bump in the grain
 size distribution around $a = 0.01\;\mu$m.
The effects of shattering in WNM and CNM on the size distribution appear
at different grain radii: While grains with $a > 0.2\;\mu$m are mainly shattered in WNM,
shattering in CNM affects grains with $a \sim 0.1\;\mu$m.
Furthermore, the effect of shattering, in particular shattering in
WNM, is large enough to determine the maximum size of grains in the
ISM.
Coagulation occurs effectively after the abundance of small grains is enhanced by shattering, and the grain size distribution is deformed to
have a bump at a larger size ($a \sim 0.03$--$0.05\;\mu$m at $t \sim
10$~Gyr) by coagulation.
We conclude that the evolution of both the total dust mass and the grain size
distribution in galaxies are related strongly to each other and 
the grain size distribution changes drastically through the
galaxy evolution.

\section*{Acknowledgments}

We thank the anonymous referee for many suggestions, which were
useful to improve the quality and the clarity of this paper.
We are grateful to Akio K.\ Inoue, Takashi Kozasa, 
Daisuke Yamasawa, and Asao Habe for fruitful discussions,
to Satoshi Okuzumi and Hiroshi Kobayashi for helpful discussions on the
process of grain{--}grain collision, to Lars Mattsson for his
comments which improve the presentation of this paper and to Jennifer
M. Stone for checking the English.
RSA acknowledges the hospitality of the members in Institute of
Astronomy and Astrophysics, Academia Sinica during his stay.
RSA has been supported from the Grant-in-Aid
for JSPS Research under Grant No.\ 23-5514.
RSA and TTT have been also partially supported from the Grand-in-Aid for the Global 
COE Program ``Quest for Fundamental Principles in the Universe: from 
Particles to the Solar System and the Cosmos'' from the Ministry of
 Education, Culture, Sports, Science and Technology (MEXT) of Japan.
TTT have been supported by the Grant-in-Aid for the Scientific 
Research Fund (TTT: 23340046, 24111707) commissioned by the MEXT and by
the Strategic Young Researcher Overseas Visits Program for Accelerating
Brain Circulation commissioned by the JSPS (R2405).
HH is supported by NSC grant 99-2112-M-001-006-MY3.
T.N. has been supported by World Premier International Research Center
Initiative (WPI Initiative), MEXT, Japan, and by the Grant-in-Aid for
Scientific Research of the JSPS (22684004, 23224004).

\appendix

\section{Examination of parameter dependence}
In this Appendix, we show dust evolution models with parameters
different from the values adopted in the main text.

\subsection{The Schmidt law index $n = 1.5$}
\label{app:schmidt}

\begin{figure*}
\centering\includegraphics[width=0.45\textwidth]{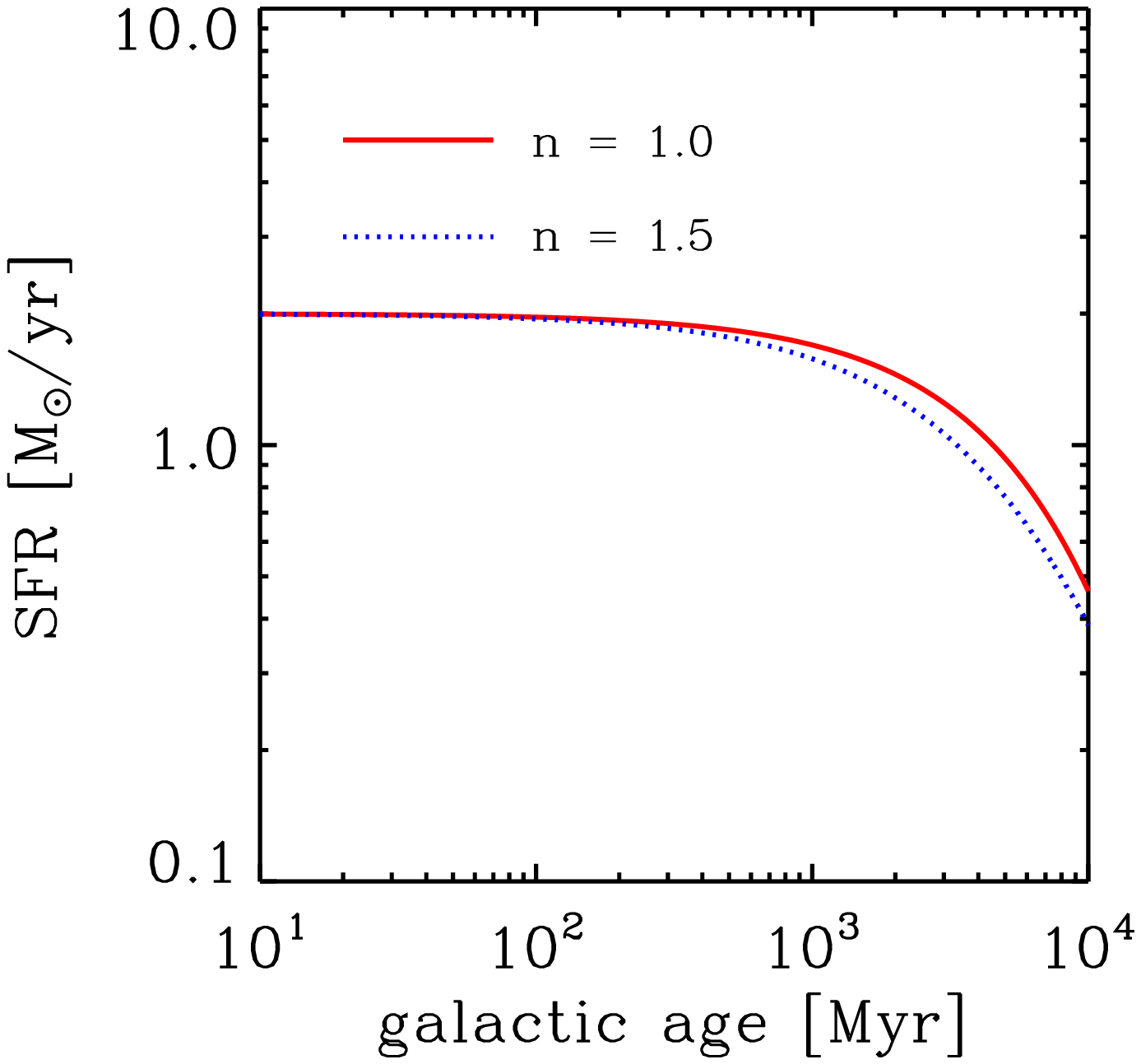}
\includegraphics[width=0.45\textwidth]{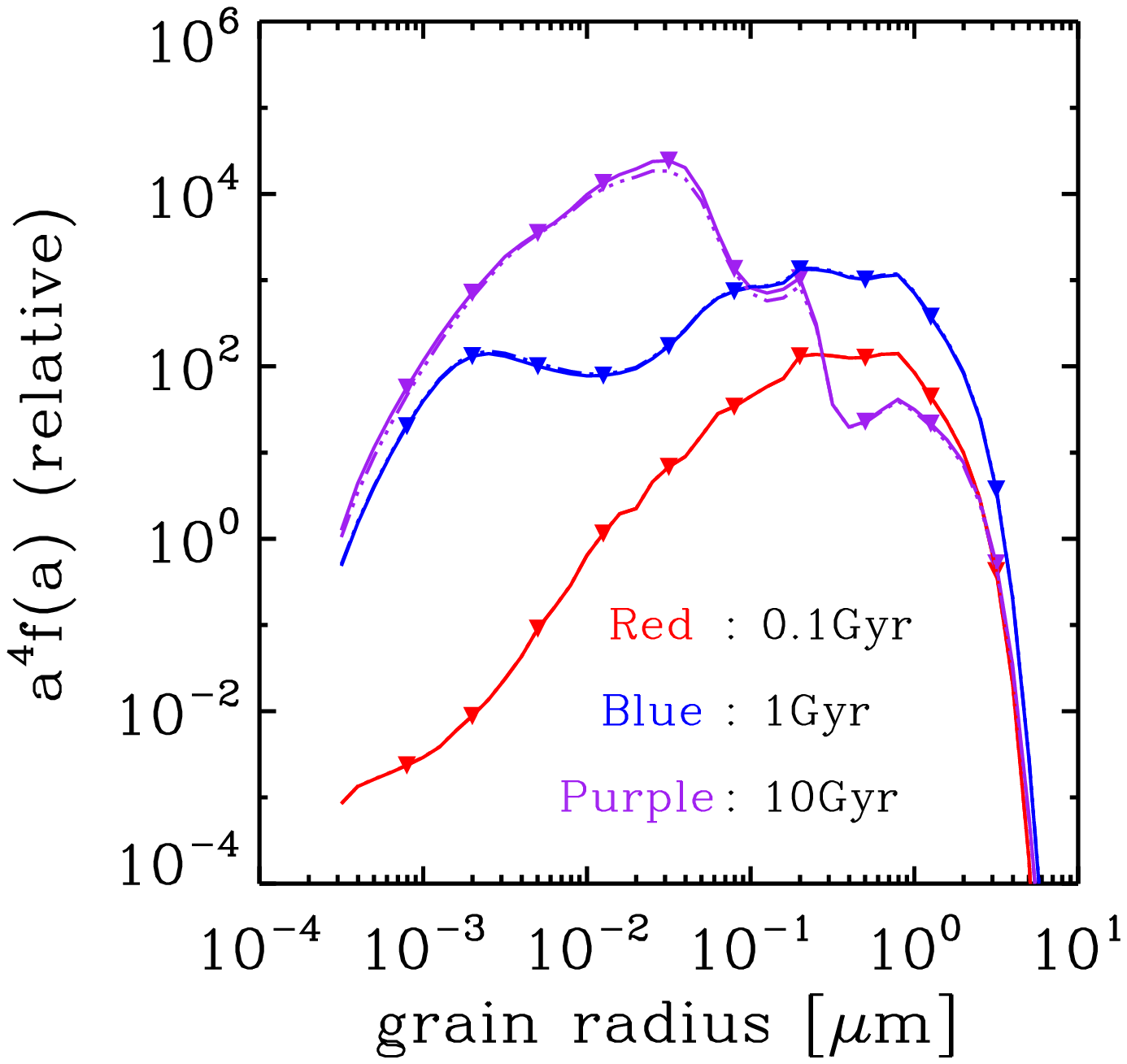}
\caption{Left panel: the star formation history with $n = 1.0$ (solid
 line) and $1.5$ (dotted line). Right panel: grain size distribution
 with $n = 1.0$ (triple-dot-dashed line) and $1.5$ (solid line with triangles).
 We adopted $\tau_{\rm SF} = 5$~Gyr, $n_{\rm SN} = 1.0\;{\rm cm}^{-3}$, and $\eta_{\rm WNM} = \eta_{\rm CNM} = 0.5$ in these plots. 
Note that the red and blue triple-dot-dashed lines overlap with the red
 and blue solid lines with triangles.
}
\label{fig:index}
\end{figure*}
In Fig.~\ref{fig:index}, 
we show star formation history (SFH) and the evolution of the grain size
distribution with the Schmidt law index~$n = 1.0$ and $1.5$. 
{}To compute the SFH and grain size distribution by using star formation rate with the Schmidt index $n = 1.5$, 
the SFR with the Schmidt law index~$n = 1.5$~(${\rm SFR}_{1.5}$) is expressed as 
\begin{equation}
{\rm SFR}_{1.5}(t) = \frac{M^{1.5}_{\rm ISM}(t)}{\nu_{1.5}},
\end{equation}
where $\nu_{1.5}$ is a constant.
We define the value of $\nu_{1.5}$ so that it satisfies the following
equation at $t = 0$:
\begin{equation}
\frac{M_{\rm ISM}(t)}{\mbox{SFR}_{1.5}(t)} = \tau_{\rm SF}.
\end{equation}
This is set to compare it with $\tau_{\rm SF}$ for $n = 1$ easily.
Thus, we obtain
\begin{equation}
\nu_{1.5} = \tau_{\rm SF}M^{0.5}_{\rm tot}.
\end{equation}
From Fig.~{\ref{fig:index}}, 
we find that the results are not
significantly different between the cases
with the $n = 1.0$ and $1.5$.

\subsection{The index of the Salpeter IMF $q = 1.35$}
\label{app:imf}

\begin{figure*}
\centering\includegraphics[width=0.45\textwidth]{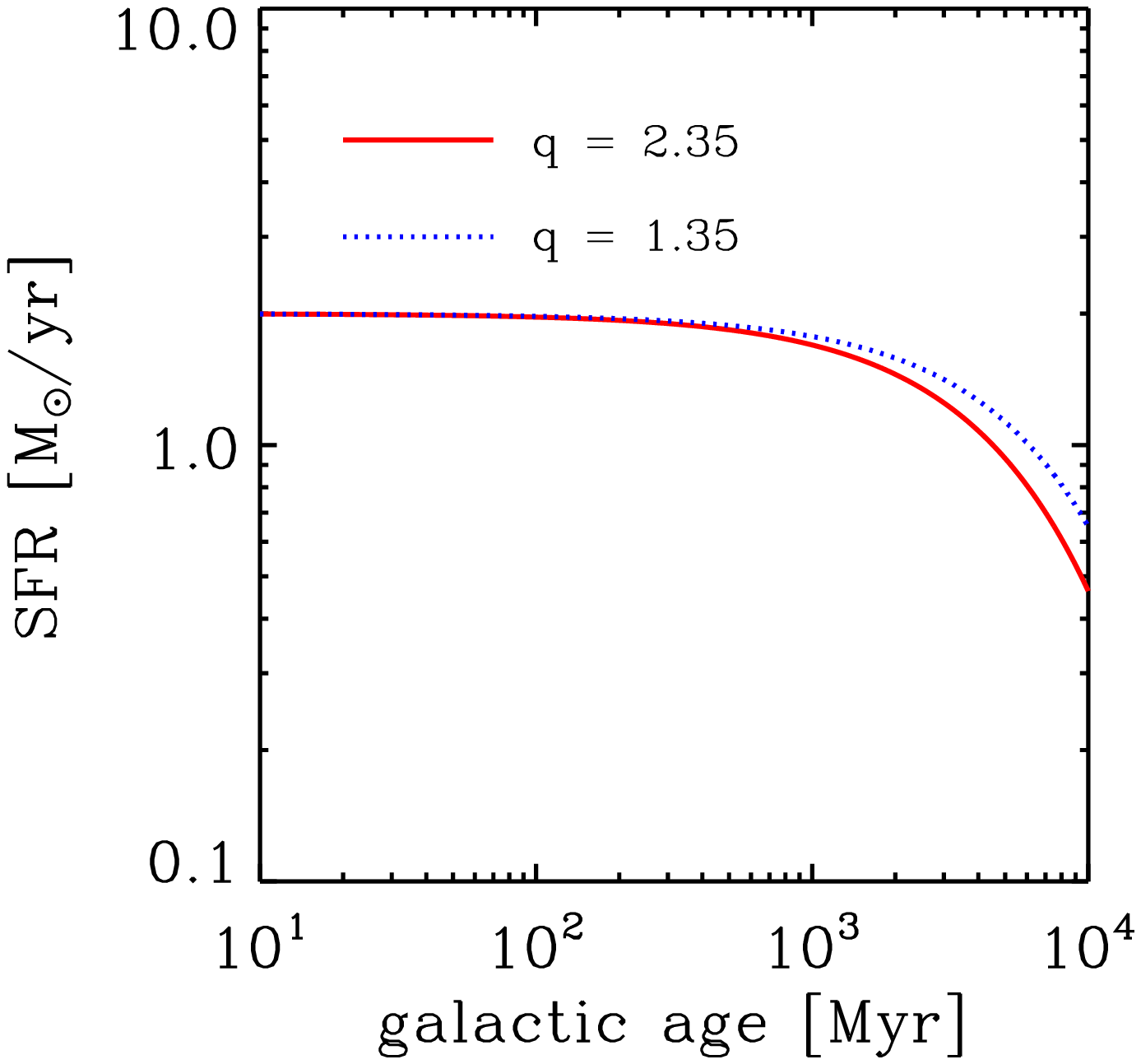}
\includegraphics[width=0.45\textwidth]{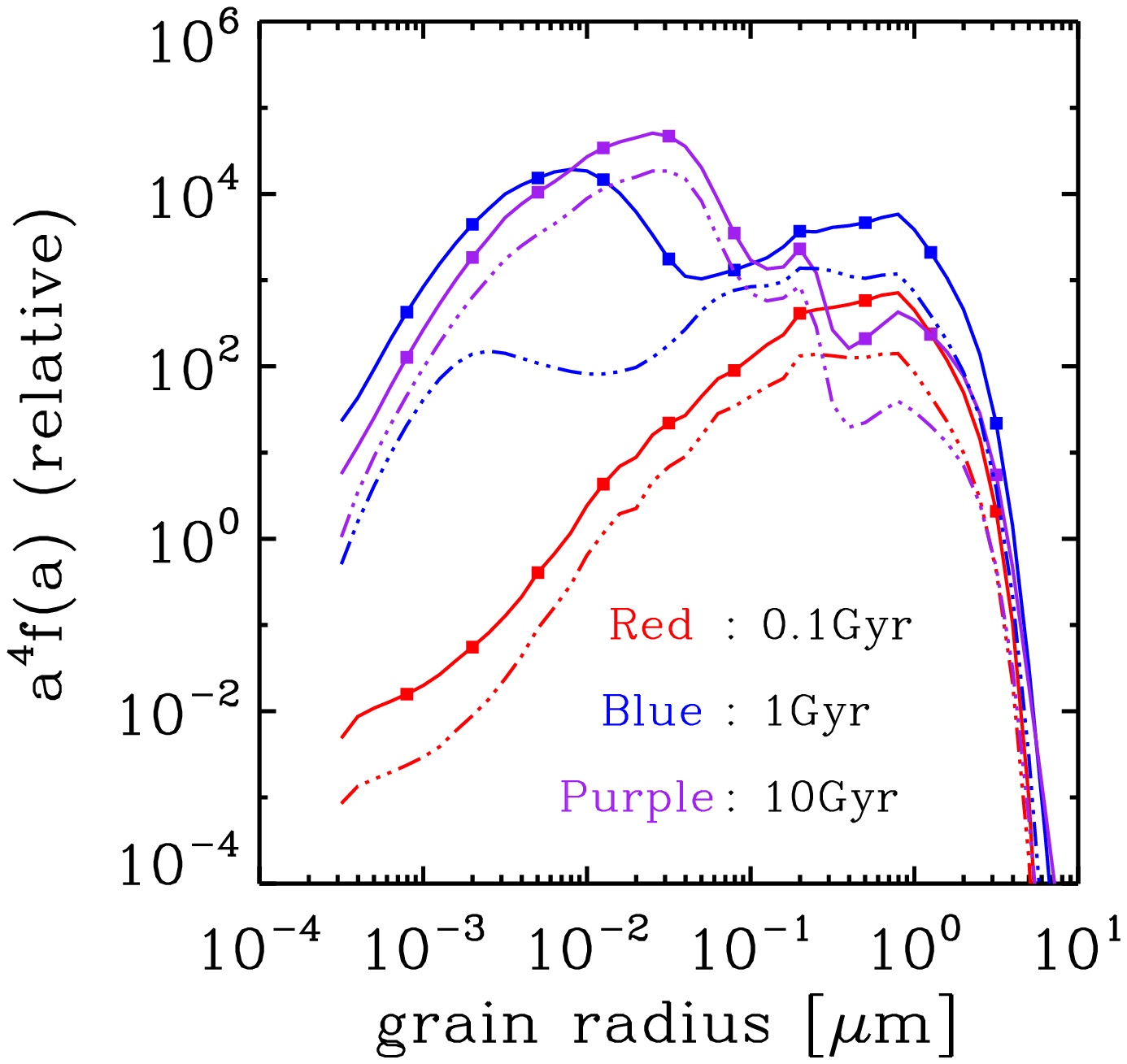}
\caption{Left panel: the star formation history with $q = 2.35$
 (solid line) and $1.35$ (dotted line). Right panel: grain size distribution
 with $q = 2.35$ (triple-dot-dashed lines) and $1.35$ (solid lines with
 filled squares).
 We adopted $\tau_{\rm SF} = 5$~Gyr, $n_{\rm SN} = 1.0\;{\rm cm}^{-3}$ and $\eta_{\rm WNM} = \eta_{\rm CNM} = 0.5$ in these plots. 
}
\label{fig:imfindex}
\end{figure*}
Figure \ref{fig:imfindex} shows the SFH and the evolution of the grain
size distribution with the power-law index of the Salpeter IMF $q = 1.35$ and
$2.35$ (fiducial value in this paper). 
We observe that SFRs are almost the same, but
the grain size distributions are different.
If $q$ is small, that is, a large number of SNe~II are produced, the
abundance of dust and metals increase earlier than the case with large $q$.
As a result, the dust amount of each size of grains [the values of $a^4f(a)$]
with $q = 1.35$ is larger than the case with $q = 2.35$.
The dust processes in the ISM also become
effective earlier because of the larger dust abundance.
However, we find that the trend of the evolution of the grain size distribution (at
early phases, stars are dominant sources of dust, as time passes, the
processes in the ISM become important) does not change.

\section{Timescale of the change of $f(a)/f(a)_{\rm star}$}
\label{app:timescale}

In Section \ref{sec:discussion}, we found that the timescale of the
change of $f(a)/f(a)_{\rm star}$ of all sizes of grains depends on star
formation timescale, and the change are due to coagulation for small grains and
shattering for large grains. 
Since both of shattering and coagulation are collisional processes,
the timescales scale with the grain abundance in the same way.
In order to evaluate the dependence on the star formation timescale, we
compare the contributions of stars and shattering.

First, we consider the stellar contribution [Eq.~(\ref{eq:stardust})].
If $D$ is defined as
\begin{equation}
D \equiv \int^{\infty}_{0}\int^{100\;{\rm M}_{\odot}}_{m_{\rm cut}(t)}
 \Delta m_{\rm d}(m, Z(t-\tau_m),a)\phi(m) {\rm d}m {\rm d}a,
\end{equation}
with Eq.~(\ref{eq:sfr}), the stellar contribution can be approximated as
\begin{equation}
\left. \frac{{\rm d}M_{\rm d}}{{\rm d}t}\right|_{\rm star} \simeq D
 \frac{M_{\rm ISM}}{\tau_{\rm SF}}. 
\label{appeq:star}
\end{equation}

Then, we consider the timescale of shattering,
$\tau_{\rm shat}$.
Since shattering is a collisional process, $\tau_{\rm shat}$ can be
represented as
\begin{equation}
\tau_{\rm shat} \simeq \frac{1}{\pi \langle a^2 \rangle v n_{\rm
 grains}},
\label{appeq:shattime}
\end{equation}
where $\langle a^2 \rangle$ is the $2$nd moment of a grain size $a$, $v$ is the relative
velocity of grains, and $n_{\rm grain}$ is the number density of grains,
which is given by
\begin{equation}
\frac{4}{3}\pi \langle a^3 \rangle s n_{\rm grain} \sim \mu n_{\rm H, shat} m_{\rm H}
 \frac{M_{\rm d}}{M_{\rm ISM}},
\label{appeq:numgra}
\end{equation}
where $\langle a^3 \rangle$ is the $3$rd moment of a grain size, $s$ is
the bulk density of dust grains, $n_{\rm H, shat}$ is the hydrogen
number density in the region where shattering occurs, and $m_{\rm H}$ is the mass of the hydrogen atom.
We assume the contribution of shattering to the amount of dust grains as
$M_{\rm d}/{\rm \tau_{\rm shat}}$,
and comparing this equation with Eq.~(\ref{appeq:star}), we obtain the
relation between shattering timescale and star formation timescale,
\begin{equation}
\tau_{\rm shat} \simeq \tau_{\rm SF} \frac{M_{\rm d}}{M_{\rm ISM}}
 \frac{1}{D}.
\label{appeq:shatstar}
\end{equation}
In addition, by substituting Eqs.~(\ref{appeq:shattime}) and
(\ref{appeq:numgra}) into Eq.~(\ref{appeq:shatstar}), we obtain
\begin{equation}
\tau_{\rm shat} \simeq \sqrt{\frac{\frac{4}{3}\pi \langle a^3 \rangle s}{\pi
 \langle a^2 \rangle v \mu m_{\rm H} n_{\rm H, shat} D}} \tau^{1/2}_{\rm SF}.
\end{equation}
To evaluate this value, we adopt $s = 3.0\;{\rm g}\;{\rm cm}^{-3}$, $v =
20\;{\rm km}\;{\rm s}^{-1}$ and $n_{\rm H, shat} = 0.3\;{\rm cm}^{-3}$
(WNM) as a representative value.
Also, from our calculation, $D \simeq 10^{-3}$, and $\langle a^3
\rangle/\langle a^2 \rangle \simeq 10^{-5}\;{\rm cm}$ for dust grains
produced by stars.
Then, we finally obtain
\begin{equation}
\tau_{\rm shat} \sim 1 \left(\frac{\tau_{\rm
			     SF}}{\mbox{Gyr}}\right)^{\frac{1}{2}}\;[{\rm
Gyr}].
\end{equation}
Thus, we conclude that the timescale of shattering, that is, the timescale of the change of
$f(a)/f(a)_{\rm star}$, is proportional to $\tau^{1/2}_{\rm SF}$.

\bsp

\label{lastpage}

\end{document}